\documentclass[a4paper,11pt]{article}
\pdfoutput=1 
\usepackage{jheppub} 

\usepackage{amsmath,amsfonts}
\usepackage{url}
\usepackage{graphicx}
\usepackage{amssymb}
\usepackage{epstopdf}
\usepackage{slashed}
\usepackage{feynmp}
\usepackage{xcolor}
\usepackage{slashbox}
\usepackage{diagbox}
\usepackage{pict2e}
\usepackage{lscape}
\usepackage{subfig}
\usepackage{ulem}
\newcommand{\OO}{\ensuremath{\mathcal{O}}}
\newcommand{\pdp}{\ensuremath{\phi^\dagger\phi}}
\renewcommand{\phi}{\ensuremath{\varphi}}

\newcommand{\sss}{\scriptscriptstyle}
\newcommand{\sst}{\scriptstyle}
\newcommand{\Op}[1]{\OO_{\sss #1}}
\newcommand{\Cp}[1]{c_{\sss #1}}

\newcommand{\bpm}{\begin{pmatrix}}      
\newcommand{\epm}{\end{pmatrix}}

 \def\lra#1{\overset{\text{\scriptsize$\leftrightarrow$}}{#1}}
\title{\boldmath Single-top associated production with a $Z$ or $H$ boson at the LHC: the SMEFT interpretation}
\author[a]{Celine Degrande,}
\author[b]{Fabio Maltoni,}
\author[b]{Ken Mimasu,}
\author[a]{Eleni Vryonidou,}
\author[c]{Cen Zhang}

\affiliation[a]{CERN, Theoretical Physics Department, Geneva 23 CH-1211, Switzerland}
\affiliation[b]{Centre for Cosmology, Particle Physics and Phenomenology (CP3),\\
	Universit\'e catholique de Louvain, B-1348 Louvain-la-Neuve, Belgium} 
\affiliation[c]{Institute of High Energy Physics, Chinese Academy of Sciences, Beijing 100049, China}

\emailAdd{celine.degrande@cern.ch}
\emailAdd{fabio.maltoni@uclouvain.be}
\emailAdd{ken.mimasu@uclouvain.be}
\emailAdd{eleni.vryonidou@cern.ch}
\emailAdd{cenzhang@ihep.ac.cn}

\abstract{At the LHC, top quarks can be produced singly with a sizeable rate via electroweak interactions. This process probes a limited set of top-quark electroweak couplings, {\it i.e.}, the same entering the top-quark decay, yet at higher scales and with a different sensitivity. Requiring the production of a $Z$ or $H$ boson in association with single-top significantly extends the sensitivity of this process to new physics, opening up the unique possibility of testing  top-Higgs, top-gauge, triple gauge, gauge-Higgs interactions without being dominated by QCD interactions. We consider $tZj$ and $tHj$ production at the LHC, providing predictions at next-to-leading accuracy in QCD in the framework of the standard model effective field theory, including all relevant operators up to dimension six. We perform the first complete study of the sensitivity to new interactions of these processes, highlighting the interplay and complementarity among $tj$, $tZj$ and $tHj$ in simultaneously constraining  top-quark, triple gauge, and gauge-Higgs interactions in the current and future runs at the LHC. }

\begin{document}
\maketitle
\flushbottom

\section{Introduction}

The study of the top-quark, gauge and Higgs boson interactions is one of the main goals of the exploration of the TeV scale at colliders. The golden era of precision physics at the LHC started after the discovery of the Higgs boson in Run I and a coordinated theoretical and experimental effort is ongoing to detect deviations and/or constrain new physics with sensitivities that go up to the multi-TeV scales. A powerful and general framework to analyse and parametrise deviations from the Standard Model (SM) predictions is the one of SM Effective Field Theory (SMEFT) \cite{Weinberg:1978kz,Buchmuller:1985jz,Leung:1984ni}, where the SM is augmented by a set of higher-dimension operators
\begin{equation} 
	\mathcal{L}_\mathrm{SMEFT}=\mathcal{L}_\mathrm{SM}+
	\sum_i\frac{C_{i}}{\Lambda^2}\OO_{i}+\mathcal{O}(\Lambda^{-4}).
\label{eq:smeft}
\end{equation}
Within the SMEFT, predictions can be systematically improved by computing higher-order corrections. Significant progress in this direction has been achieved in both the top-quark \cite{Zhang:2013xya,Zhang:2014rja,Degrande:2014tta,Franzosi:2015osa,Zhang:2016omx,Bylund:2016phk,Maltoni:2016yxb,Rontsch:2014cca,Rontsch:2015una} and Higgs sectors \cite{Hartmann:2015oia,Ghezzi:2015vva,Hartmann:2015aia,Gauld:2015lmb,Mimasu:2015nqa,Degrande:2016dqg}. 

Among the least known interactions between the heaviest particles of the standard model are the neutral gauge and Higgs top-quark interactions. These interactions can be probed directly for the first time at the LHC through the associated production of a Higgs, $Z$ or $\gamma$ with a top-quark pair. In this case the leading production mechanisms are through QCD interactions (at order $\alpha_S^2$ at the Born level) and both theoretical studies and experimental ones exist that establish the present and future sensitivities ~\cite{Degrande:2012gr,Rontsch:2014cca,Rontsch:2015una,Bylund:2016phk,Schulze:2016qas,Maltoni:2016yxb} to new couplings as parametrised in the SMEFT. An intrinsic limitation of this strategy is the fact that a plethora of operators enter these processes some of which are of QCD nature or involve four fermions. Therefore they need to be constrained very well (through, for example, $t \bar t$ production) before being able to access the electroweak ones.

A promising alternative, discussed in this work, is to consider the corresponding set of associated production processes of neutral heavy bosons with a single top.  At the LHC top quarks can be produced singly via electroweak interactions, the leading process being $t$-channel production ($tj$), $qb \to q't$, which features a total single top and anti-top rate which is about $220$ pb at $\sqrt{S}=13$ TeV, {\it i.e.}, one fourth of strong $t \bar t$ production. The cross section probes a limited set of top-quark electroweak couplings, {\it i.e.}, at leading order, two four-quark interactions and three operators which induce a modification of the top electroweak couplings. Considering also the top decay one can additionally probe top-quark four-fermion operators involving leptons. Requiring a  $Z$ or  a $H$ boson in association with single-top significantly extends the sensitivity of $tj$, opening up the rather unique possibility of accessing  top-Higgs, top-gauge, triple gauge, gauge-Higgs interactions in the same final state.\footnote{We have explicitly verified that $t\gamma j$ production displays, in fact, similar sensitivities to  new neutral gauge and top-quark interactions as $tZj$ and that the corresponding predictions at the LHC  can be automatically obtained at NLO in QCD in our framework. As no dedicated experimental analysis of this process is available yet, we defer a detailed study to the future.}   The fact that these processes can play an important role in the search for new neutral top-quark interactions has been already noted at the theory level~\cite{Maltoni:2001hu,Biswas:2012bd,Farina:2012xp,Demartin:2015uha} (even though not yet analysed in the context of the SMEFT) and motivated experimental activities, such as the measurements of the associated production of a $Z$ with a single top quark by ATLAS \cite{Aaboud:2017ylb} and CMS \cite{CMS-PAS-TOP-16-020,Sirunyan:2017nbr} at 13 TeV, as well as the searches for $tHj$ production, which are also underway \cite{CMS-PAS-HIG-16-019,CMS-PAS-HIG-17-005}. In addition, asking for just one top-quark (or anti-top-quark) in the final state implies no QCD interactions at the leading order (LO) and therefore makes this class of processes `purely' electroweak with two important consequences. First, SM QCD corrections are typically small and under control. Second,  dim-6 modifications of QCD interactions enter only at NLO with a weak sensitivity that does not spoil that of the EW couplings. 
\begin{figure}[t]
    \centering
    \includegraphics[width=0.8\textwidth]{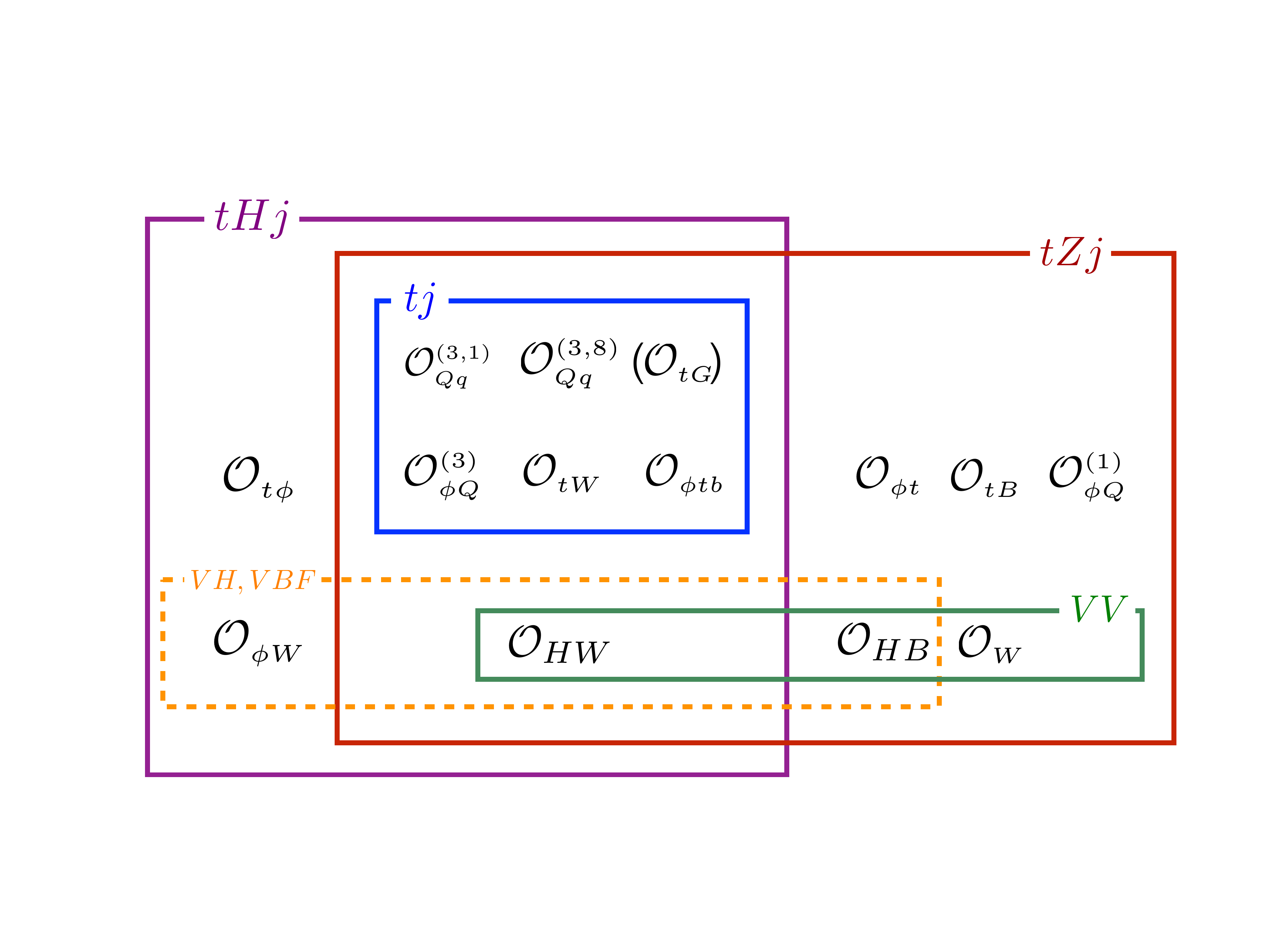}
    \vspace*{-1cm}
    \caption{\label{fig:eftdiagram}Schematic representation of the interplay between operators and processes, focussing on single-top production and associated channels.  Six (five at LO and one at NLO in QCD) operators enter single-top production ($tj$, blue square), and are therefore also present  in $Z$ boson  ($tZj$, red square) and in Higgs ($tHj$, purple square) associated production. Operators exist that contribute to either $tZj$ or $tHj$ and also to both processes without contributing to $tj$. The operators entering in diboson ($VV$) production are a subset (green square) of those contributing to $tZj$, while some of the operators contributing to Higgs associated production ($VH$) and Vector Boson Fusion (VBF, orange dashed square) are shared between $tHj$ and $tZj$.} 
\end{figure}

In this work, we consider the $t$-channel $tZj$ and $tHj$ production at the LHC, providing predictions at  NLO accuracy in QCD in the general framework of the SMEFT, including all relevant operators up to dimension six.  This is the first time NLO in QCD corrections are calculated for processes that involves all possible types of dim-6 operators, {\it i.e.} bosonic, two-fermion and four-fermion ones in a fully automatic way.  We perform a complete study of the sensitivity to new interactions of these processes, highlighting the interplay and complementarity among $tj$, $tZj$ and $tHj$ in simultaneously constraining  top-quark, triple gauge, and gauge-Higgs interactions in the current and future runs at the LHC (see Fig. \ref{fig:eftdiagram}). 

We first study the energy dependence of relevant $2\to2$ sub-amplitudes to identify the set of operators that may induce deviations in each process and characterise the expected energy growth in each case. We then compute the complete dependence of the inclusive rates on these operators at NLO in QCD, including estimates of the scale uncertainty due to the running of the Wilson coefficients where applicable. Our approach is based on the {\sc MadGraph5\_aMC@NLO} ({\sc MG5\_aMC}) framework \cite{Alwall:2014hca}, and is part of the ongoing efforts of automating NLO SMEFT simulations for colliders~\cite{Zhang:2016snc}. Using these results, we perform sensitivity studies of current and future inclusive measurements of the two processes, contrasting them with existing limits on the operators of interest. Finally, we present differential distributions for a number of selected benchmark values of the Wilson coefficients inspired by current limits, highlighting the possibility of large deviations in the high energy regime of both processes.

This paper is organised as follows. In Section~\ref{section:Operators} we establish the notation and the conventions, we identify the set of operators entering $tj$, $tZj$ and $tHj$ and we establish which ones can lead to an energy growth. In Section~\ref{section:constraints} a summary of the current constraints available on the Wilson coefficients of the corresponding operators is given. In Section~\ref{sec:setup} results for total cross sections as well as distributions are presented, operator by operator and the prospects of using $tZj$ and $tHj$ to constrain new interactions are discussed. The last section presents our conclusions and the outlook.

\section{Top-quark, electroweak and Higgs operators in the SMEFT} 
\label{section:Operators}
The processes that we are studying lie at the heart of the electroweak symmetry breaking sector of the SM. They involve combinations of interactions between the Higgs boson and the particles to which it is most strongly coupled: the top quark and the EW gauge bosons. The measurement of these processes is therefore a crucial test of the nature of EW symmetry breaking in the SM and any observed deviations could reveal hints about the physics that lies beyond. 

We adopt the SMEFT framework to parametrise the deviations of the interactions in question from SM expectations. Dim-6 operators suppressed by a scale, $\Lambda$, are added to the SM Lagrangian as in Eq.~(\ref{eq:smeft}).
Specifically, we employ the Warsaw basis~\cite{Grzadkowski:2010es} dim-6
operators relevant for the $tHj$ and $tZj$  processes. To this end, it is
convenient to work in the limit of Minimal Flavour Violation
(MFV)~\cite{DAmbrosio:2002vsn}, in which it is assumed that the only sources of
departure from the global $U(3)^5$ flavour symmetry of the SM arise from the
Yukawa couplings.  By assuming a diagonal CKM matrix and only keeping operators
with coefficients proportional to the third generation Yukawas, we retain all operators in the top-quark sector, as well as all the light-fermion operators that are flavour-universal
\cite{AguilarSaavedra:2018nen}. Assuming in addition $CP$ conservation, we are
left with a well-defined set of operators that can directly contribute to the
processes, summarised in Table~\ref{tab:operators} where all Yukawa and gauge
coupling factors are assumed to be absorbed in the operator coefficients. We
adopt the following definitions and conventions:
\begin{align}
  \phi^\dag {\overleftrightarrow D}_\mu \phi&=\phi^\dag D^\mu\phi-(D_\mu\phi)^\dag\phi\nonumber\\
  \phi^\dag \tau_{\sss K} {\overleftrightarrow D}^\mu \phi&=
  \phi^\dag \tau_{\sss K}D^\mu\phi-(D^\mu\phi)^\dag \tau_{\sss K}\phi\nonumber \\
  W^{\sss K}_{\mu\nu} &= \partial_\mu W^{\sss K}_\nu 
  - \partial_\nu W^{\sss K}_\mu 
  + g \epsilon_{\sss IJ}{}^{\sss K} \ W^{\sss I}_\mu W^{\sss J}_\nu \nonumber\\
  B_{\mu\nu} &= \partial_\mu B_\nu - \partial_\nu B_\mu \\
  D_\rho W^{\sss K}_{\mu\nu} &= \partial_\rho W^{\sss K}_{\mu\nu} + g \epsilon_{\sss IJ}^{\sss K} W^{\sss I}_\rho W_{\mu\nu}^{\sss J}\nonumber \\
  D_\mu\phi =& (\partial_\mu -  i g \frac{\tau_{\sss K}}{2} W_\mu^{\sss K} - i\frac12 g^\prime B_\mu)\phi,\nonumber
\end{align}
where $\tau_I$ are the Pauli matrices.

\begin{table}[h!]
    \centering
    \renewcommand{\arraystretch}{1.4}
{  \centering
\begin{tabular}{|ll|ll|}
    \hline
     $\Op{W}$&
     $\varepsilon_{\sss IJK}\,W^{\sss I}_{\mu\nu}\,
                             {W^{{\sss J},}}^{\nu\rho}\,
                             {W^{{\sss K},}}^{\mu}_{\rho}$&
     $\Op{\phi Q}^{\sss(3)}$&
     $i\big(\phi^\dagger\lra{D}_\mu\,\tau_{\sss I}\phi\big)
     \big(\bar{Q}\,\gamma^\mu\,\tau^{\sss I}Q\big)
     + \text{h.c.}$         
      \tabularnewline
     $\Op{\phi W}$&
     $\left(\pdp-\tfrac{v^2}{2}\right)W^{\mu\nu}_{\sss I}\,
                                    W_{\mu\nu}^{\sss I}$&
     $\Op{\phi Q}^{\sss(1)}$&
     $i\big(\phi^\dagger\lra{D}_\mu\,\phi\big)
     \big(\bar{Q}\,\gamma^\mu\,Q\big)
     + \text{h.c.}$
     \tabularnewline
     $\Op{\phi WB}$&
     $(\phi^\dagger \tau_{\sss I}\phi)\,B^{\mu\nu}W_{\mu\nu}^{\sss I}\,$&
     $\Op{\phi t}$&
     $i\big(\phi^\dagger\lra{D}_\mu\,\phi\big)
     \big(\bar{t}\,\gamma^\mu\,t\big)
     + \text{h.c.}$
     \tabularnewline
     $\Op{\phi D}$&
     $(\phi^\dagger D^\mu\phi)^\dagger(\phi^\dagger D_\mu\phi)$&
     $\Op{\phi tb}$&
     $i\big(\tilde{\phi}\,{D}_\mu\,\phi\big)
     \big(\bar{t}\,\gamma^\mu\,b\big)
     + \text{h.c.}$
     \tabularnewline
     $\Op{\phi \square}$&
     $(\varphi^\dagger\varphi)\square(\varphi^\dagger\varphi)$ &
     $\Op{\phi q}^{\sss(1)}$&
     $i\big(\phi^\dagger\lra{D}_\mu\,\phi\big)
     \big(\bar{q}_i\,\gamma^\mu\,q_i\big)
     + \text{h.c.}$\tabularnewline
     \cline{1-2}
     $\Op{t\phi}$&
     $\left(\pdp-\tfrac{v^2}{2}\right)
     \bar{Q}\,t\,\tilde{\phi} + \text{h.c.}$&
     $\Op{\phi q}^{\sss(3)}$&
     $i\big(\phi^\dagger\lra{D}_\mu\,\tau_{\sss I}\phi\big)
     \big(\bar{q}_i\,\gamma^\mu\,\tau^{\sss I}q_i\big)
     + \text{h.c.}$
     \tabularnewline
     $\Op{tW}$&
     $i\big(\bar{Q}\sigma^{\mu\nu}\,\tau_{\sss I}\,t\big)\,
     \tilde{\phi}\,W^I_{\mu\nu}
     + \text{h.c.}$&
     $\Op{\phi u}$&
     $i\big(\phi^\dagger\lra{D}_\mu\,\phi\big)
     \big(\bar{u}_i\,\gamma^\mu\,u_i\big)
     + \text{h.c.}$
        \tabularnewline
   \cline{3-4}
     $\Op{tB}$&
     $i\big(\bar{Q}\sigma^{\mu\nu}\,t\big)
     \,\tilde{\phi}\,B_{\mu\nu}
     + \text{h.c.}$&
          $\Op{Qq}^{\sss (3,1)}$&
    $\big(\bar{q}_i\,\gamma_\mu\,\tau_{\sss I}q_i\big)
      \big(\bar{Q}\,\gamma^\mu\,\tau^{\sss I}Q\big)$
      \tabularnewline

     $\Op{tG}$&
     $i\big(\bar{Q}\sigma^{\mu\nu}\,T_{\sss A}\,t\big)
     \,\tilde{\phi}\,G^{\sss A}_{\mu\nu}
     + \text{h.c.}$&
         $\Op{Qq}^{\sss (3,8)}$&
    $\big(\bar{q}_i\,\gamma_\mu\,\tau_{\sss I}T_{\sss A} q_i\big)
      \big(\bar{Q}\,\gamma^\mu\,\tau^{\sss I}T^{\sss A} Q\big)$
      \tabularnewline
     \hline
  \end{tabular}}

    \caption{\label{tab:operators} Dim-6 operators relevant for the $tZj$ and $tHj$ processes in the Warsaw basis. The first set corresponds to bosonic operators, then two-fermion ones, and, finally, four fermion operators.}
\end{table} 
We compute predictions for on-shell top quark, Higgs and $Z$ bosons, ignoring operators that could mediate the same decayed final state through a contact interaction such as the $\bar{t}t\bar{\ell}\ell$ four-fermion operators. This contribution is expected to be suppressed, as the experimental analyses typically apply a cut on the invariant mass of the lepton pair around the $Z$ mass. It can nevertheless be straightforwardly included as was done in \cite{Durieux:2014xla}. 
Two four-quark operators that also mediate single-top production do affect these processes. 
These operators $ \Op{Qq}^{\sss (3,1)}$ and $ \Op{Qq}^{\sss (3,8)}$ listed in Table~\ref{tab:operators} contribute at $1/\Lambda^2$ and $1/\Lambda^4$ respectively, the latter not interfering with the SM processes at LO due to colour. While the measurement of the single-top process already constrains these operators, the higher kinematic thresholds of the associated production may enhance the dependence on the Wilson coefficients. 

In addition, the following operators contribute indirectly, by affecting the
muon decay and consequently the relation between the Fermi constant and the
Higgs vacuum-expectation-value:
\begin{align}
    \Op{ll}^{\sss (3)}& = (\bar{l}_i\,\gamma_\mu\tau_{\sss I}\,l_i)
                         (\bar{l}_j\,\gamma^\mu\tau^{\sss I}\,l_j),\\
    \Op{\phi l}^{\sss (3)}& =
     i\big(\phi^\dagger\lra{D}_\mu\,\tau_{\sss I}\phi\big)
     \big(\bar{l}_{i}\,\gamma^\mu\,\tau^{\sss I}l_{i}\big)
     + \text{h.c.}\,.
     \end{align}
Some of these operators are constrained by Electroweak Precision Observables EWPO \cite{Han:2004az}.
These include the two previous operators and those involving light-fermion fields, {\it i.e.},~$\Op{\phi q}^{(1)}$,
$\Op{\phi q}^{\sss(3)}$, $\Op{\phi u}$, $\Op{\phi d}$,
$\Op{\phi l}^{\sss(3)}$, $\Op{\phi l}^{\sss(1)}$, $\Op{\phi e}$, $\Op{ll}^{\sss (3)}$,
where
\begin{flalign}
	&\Op{\phi d}=i\big(\phi^\dagger\lra{D}_\mu\,\phi\big)
	\big(\bar{d}_i\,\gamma^\mu\,d_i\big) + \text{h.c.}
	\\
	&\Op{\phi l}^{\sss(1)} =i\big(\phi^\dagger\lra{D}_\mu\,\phi\big)
     \big(\bar{l}_i\,\gamma^\mu\,l_i\big)
     + \text{h.c.}
	\\
	&\Op{\phi e}=i\big(\phi^\dagger\lra{D}_\mu\,\phi\big)
	\big(\bar{e}_i\,\gamma^\mu\,e_i\big) + \text{h.c.}\,,
\end{flalign}
as well as the operators that are often identified with the S and T parameters
\begin{align}
     \Op{\phi WB} &=  (\phi^\dagger \tau_{\sss I}\phi)\,B^{\mu\nu}W_{\mu\nu}^{\sss I}\,\text{  and  }\\
     \Op{\phi D} &=
      (\phi^\dagger D^\mu\phi)^\dagger(\phi^\dagger D_\mu\phi).
\end{align}
It is well-known that among these 10 basis operators, only 8 degrees of freedom
are tightly constrained \cite{Falkowski:2014tna}, leaving two flat directions that are
constrained only by diboson production processes.  This effect has been
discussed in the literature \cite{Grojean:2006nn,Alonso:2013hga,Brivio:2017bnu}.  These two
directions correspond exactly to the two basis-operators in the HISZ parametrisation
\cite{Hagiwara:1993ck}:
\begin{eqnarray}
\Op{HW}&=&(D^\mu\phi)^\dagger\tau_{\sss I}(D^{\nu}\phi)W^{\sss I}_{\mu\nu}\\
\Op{HB}&=&(D^\mu\phi)^\dagger(D^{\nu}\phi)B_{\mu\nu}.
\end{eqnarray}
Apart from modifying the Higgs couplings, the coefficients of these two
operators are often used to parametrise the triple-gauge-boson (TGC) couplings,
together with the coefficient of $\Op{W}$ \cite{Degrande:2012wf}. They can be
determined by di-boson and tri-boson production processes.  Since one interesting application 
of this work is to determine the sensitivity of the $tZj$ and $tHj$ processes to
TGC couplings relative to the di-boson processes, we include these two
additional operators to cover all possible Lorentz structures in TGC modifications from dim-6 SMEFT.  With this
choice we can safely exclude the 10 Warsaw basis operators that enter the EWPO
measurements.  We also neglect the operator
$(\varphi^\dagger\varphi)\square(\varphi^\dagger\varphi)$, which universally
shifts all Higgs couplings.  This operator does not lead to any different
energy-dependent behaviour, and is likely to be better constrained by other
Higgs measurements.

We briefly mention here that the complete RG structure of the SMEFT has been given in
\cite{Jenkins:2013zja,Jenkins:2013wua,Alonso:2013hga}.  In this work we will
consider the QCD induced running of the Wilson coefficients, which is relevant for our calculation,
{\it i.e.}~$\mathcal{O}(\alpha_s)$ terms with our normalisation.  The only operators 
from our set that run under QCD are $(\Op{t\phi},\Op{tW},\Op{tB})$. The mixing matrix  has a diagonal form:
\begin{equation}
	\frac{dC_i(\mu)}{d\log\mu}=\frac{\alpha_s}{\pi}\gamma_{ij}C_j(\mu),\quad
	\gamma= \left(
	\begin{array}{ccc}
		-2 & 0 & 0 \\
		0  & 2/3 & 0 \\
		0 & 0 & 2/3
	\end{array}
\right)\,.
\label{eq:rg}
\end{equation}

The chromomagnetic operator, $\Op{tG}$, also mixes into the weak dipole operators at NLO in QCD and therefore contributes to our two processes at one-loop. While this is an interesting effect, we do not expect to obtain significant additional information from $tZj$ or $tHj$ given the current constraints from top measurements and the fact that it enters at higher order in $\alpha_{\sss S}$. We nevertheless compute its contribution to our processes for completeness.

In summary, the operators to be considered in this work are:
\begin{flalign}
	\label{oplist1}
	&\mbox{Pure gauge operators (4):}\quad
	\Op{\phi W},\Op{W},\Op{HW},\Op{HB},\\
	\label{oplist2}
	&\mbox{Two-fermion top-quark operators (8):}\quad
	\Op{\phi Q}^{\sss (3)},\Op{\phi Q}^{\sss (1)},\Op{\phi
	t},\Op{tW},\Op{tB},\Op{tG},\Op{\varphi tb},\Op{t\phi},\\
	\label{oplist3}
	&\mbox{Four-fermion top-quark operators (2):}\quad
	\Op{Qq}^{\sss(3,1)},\Op{Qq}^{\sss(3,8)}.
\end{flalign}

\begin{figure}
    \centering
    \includegraphics[width=0.2\textwidth]{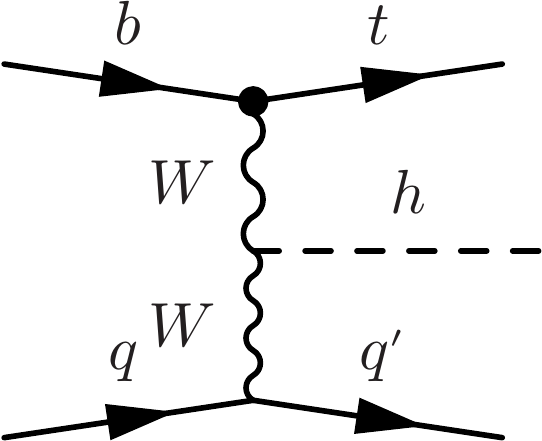}
    \includegraphics[width=0.2\textwidth]{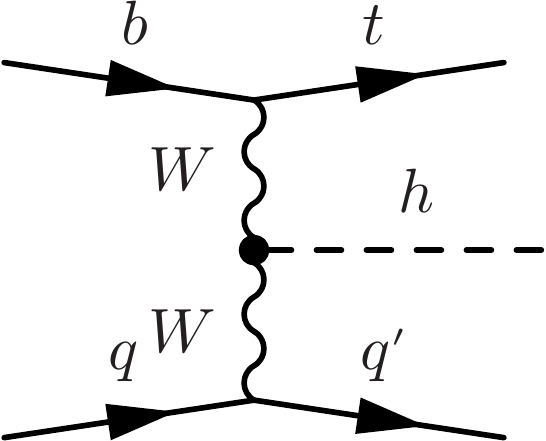}
    \includegraphics[width=0.2\textwidth]{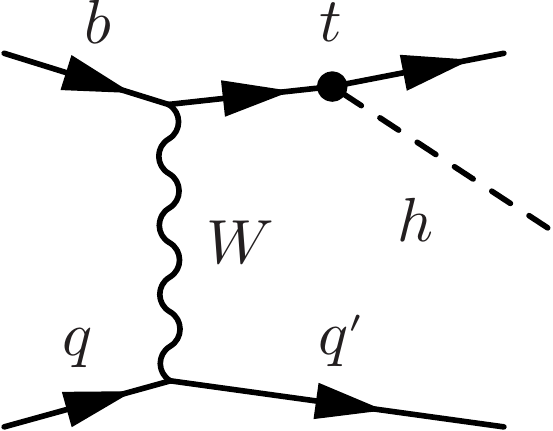}
    \includegraphics[width=0.2\textwidth]{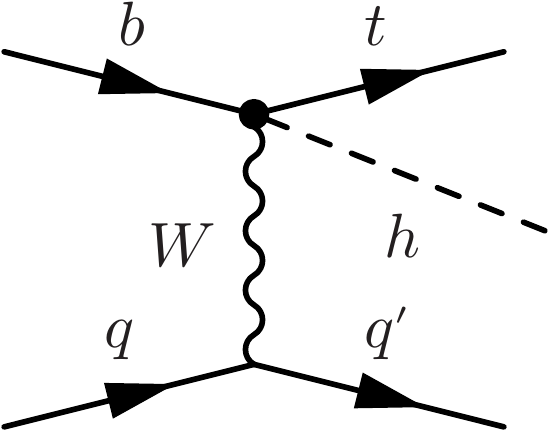}
    \caption{\label{fig:feyndiags}
    Representative LO Feynman diagrams for $tHj$ production in the SMEFT. The operator insertions (black dots) correspond to operators involving either electroweak gauge boson or third generation fermions interactions. These can modify existing SM interactions such as the top Yukawa or Higgs-$W$-$W$ interaction, induce new Lorentz structures, {\it e.g.}, with the weak dipole operators or mediate new contact interactions between fermion currents and two EW bosons. Equivalent diagrams for the $tZj$ process can be obtained by replacing the Higgs with a $Z$ boson and keeping in mind that the $Z$ boson can also couple to the light-quark line.
    }
\end{figure}

Figure \ref{fig:eftdiagram} gives a visual representation of how different operators contribute to the set of processes $tj$, $tZj$ and $tHj$, and also $VV$ and $VH$,VBF production. As mentioned already, an interesting feature of $tZj$ and $tHj$ is that they are affected by the same operators that enter $ttZ$ and $ttH$, respectively, yet they are entangled in a non-trivial way. The connection of different sectors by these two processes is required by the nature of SMEFT \cite{Corbett:2013pja,Falkowski:2015jaa,Maltoni:2016yxb} and makes these processes a unique testing ground for operators at the heart of the EW symmetry breaking sector.  Figure~\ref{fig:feyndiags} shows a selection of representative Feynman diagrams for the $tHj$ process in which the SMEFT modifications can enter.

\subsection{Energy growth and sub-amplitudes
\label{sec:energybehaviour}}
One of the characteristic ways in which anomalous interactions between SM particles manifest themselves is through the energy growth of the scattering amplitudes. An enhancement can arise through two basic mechanisms. The first is due to vertices involving higher dimension Lorentz structures, {\it i.e.}, with additional derivatives or four-fermion interactions. The second, more subtle, can come from deformations induced by operators that do not feature new Lorentz structures, yet spoil delicate unitarity cancellations that might take place in the SM amplitudes. In general, higher dimensional operators involving Higgs fields can contribute to either of these effects, given that insertions of the Higgs vacuum-expectation-value can lower the effective dimension of a higher -dimesion operator down to dim-4. A concrete example of this phenomenon can be found in $tHj$, where both the diagram featuring the top-quark Yukawa coupling and the one with the $W$-Higgs interaction (see the second and third diagrams of Fig.~\ref{fig:feyndiags}), grow linearly with energy, and yet this unitarity-violating dependence exactly cancels in the SM~\cite{Bordes:1992jy,Maltoni:2001hu,Farina:2012xp}.  The rate of this process is therefore  sensitive to the deviations in the Higgs couplings to the top quark and $W$-boson. This can be understood by factorising the process into the emission of an on-shell $W$-boson from the initial light quark, weighted by an appropriate distribution function, times the $b\,W\to t\,h$ sub-amplitude. The sub-amplitudes of the two diagrams in question, involving a longitudinally polarised $W$, both display an unacceptable energy growth which cancels in the SM limit. Similarly for the $tZj$ process, the $b\,W\to t\,Z$ sub-amplitude for longitudinally polarised gauge bosons can suffer from such behaviour away from the SM limit. We note here that whilst $tHj$ essentially always proceeds through the $b\,W\to t\,h$ sub-amplitude, the $b\,W\to t\,Z$ sub-amplitude is not the only one contributing to $tZj$ as the $Z$ can be emitted from light quark lines.

In the framework of SMEFT, the high energy behaviour of the $2\to2$ sub-amplitudes as a function of the Wilson coefficients for $b\,W\to t\,h$ and $b\,W\to t\,Z$ for the operators in Eqs.~(\ref{oplist1}-\ref{oplist3}) are shown in Tables~\ref{tab:bwth} and~\ref{tab:bwtz} respectively. One can see that the energy growth due to higher dimensional operators can arise from sources other than additional derivatives, {\it i.e}, from the top Yukawa and Higgs-fermion current operators. The former operator only modifies the SM Higgs-top coupling and its energy dependence is a manifestation of the previously discussed unitarity violating behaviour. The latter operators both modify the SM gauge boson coupling to fermions and induce a $f\bar{f}VH$ contact term as in the last diagram of Fig.~\ref{fig:feyndiags}. 
A more complete study of the full set of top \& EW $2\to2$ subamplitudes in the SMEFT and associated LHC processes is on going and will appear in future work. In the meantime we keep these tables for reference and to help put into context the energy dependence of the results of our predictions. Overall, the possible energy enhancements in these channels suggest that, although $tZj$ and particularly $tHj$ are rare processes in the SM, such behaviour might nevertheless lead to interesting constraints on the operators studied, especially at differential level.

\begin{table}[h!]
    \centering
   \renewcommand{\arraystretch}{1.4}
\begin{center}
\begin{tabular}{|c|c|cccc|c|}
\hline
${\sst\lambda_b}$, ${\sst\lambda_{\sss W}}$, ${\sst\lambda_t}$ & SM & $\Op{t \phi}$ & $\Op{\phi Q}^{\sss (3)}$ & $\Op{\phi W}$ & $\Op{tW}$ & $\Op{HW}$\tabularnewline
\hline
$-, 0, -$& 
$s^0$ & $s^0$ & ${\sqrt{s (s+t)}}$ & $s^0$ & $s^0$& $\sqrt{s (s+t)}$\tabularnewline
$-, 0, +$& 
$\frac{1}{\sqrt{s}}$ & $ {\scriptstyle m_t\sqrt{-t} }$ & ${\scriptstyle  m_t \sqrt{-t}}$  & $\frac{1}{\sqrt{s}}$ & $\frac{ m_W s }{\sqrt{-t}}$ & $\frac{1}{\sqrt{s}}$\tabularnewline
$-, -, -$& 
$\frac{1}{\sqrt{s}}$ & $\frac{1}{\sqrt{s}}$ & $ {\scriptstyle m_W\sqrt{-t}}$ & $\frac{m_W s }{\sqrt{-t}}$ & $ {\scriptstyle m_t \sqrt{-t}} $ & $\frac{m_W (s +t)}{\sqrt{-t}}$\tabularnewline
$-, -, +$& 
$\frac{1}{s}$ & $s^0$ & $s^0$& $-$ & ${\scriptstyle \sqrt{s (s+t)}}$ & $\frac{1}{s}$\tabularnewline
$-, +, -$& 
$\frac{1}{\sqrt{s}}$ & $-$ & $\frac{1}{\sqrt{s}}$ & $\frac{m_W (s+t)}{\sqrt{-t}}$ & $\frac{1}{\sqrt{s}}$ & $\frac{m_W (s +t)}{\sqrt{-t}}$\tabularnewline
$-, +, +$& $s^0$ & $-$ & $s^0$& $s^0$ & $s^0$ & $\frac{1}{s}$\tabularnewline
\hline
\end{tabular}

\renewcommand{\arraystretch}{1.}
\vspace{0.5cm}

$\Op{\phi tb},\,\lambda_b=+$\\
\begin{tabular}{|c|ccc|}
\hline
\diagbox[height=0.8cm,innerwidth=0.8cm,font=\footnotesize]{${\sst\lambda_t}$}{${\sst\lambda_{\sss W}}$}&0&$+$&$-$\tabularnewline
\hline
\renewcommand{\arraystretch}{1.4}
$+$& ${\scriptstyle \sqrt{s (s+t)}}$ & ${\scriptstyle  m_W \sqrt{-t} }$ &  $\frac{1}{\sqrt{s}}$  \tabularnewline
$-$& ${\scriptstyle m_t \sqrt{-t}} $ & $s^0$   & $s^0$ \tabularnewline
\hline
\end{tabular}
%
\end{center}

    \caption{\label{tab:bwth}
    Energy growth of the helicity amplitudes in the $b\,W\to t\,H$ subamplitude in 
    the high energy limit, $s , -t \gg v$ with $s/-t$ constant. 
    Schematic energy growths for the SM are also shown and $s^0$ denotes constant 
    behaviour with energy. The RHCC operator ($\Op{\phi tb}$) contributions are collected 
    separately due to the fact that it is the only operator that can yield right 
    handed $b$-quark configurations in the 5-flavour scheme.
    }
\end{table} 

\section{Constraints on dim-6 operators}
\label{section:constraints}
In order to examine the sensitivity of our processes to SMEFT operators
we first consider the current limits on the dim-6 operators of interest. 
We briefly summarise the current constraints in Table~\ref{tab:constraints}.  Firstly, all top-quark operators can be constrained using collider measurements. For example, the
TopFitter collaboration has performed a global fit (excluding $\Op{\phi tb}$) at LO using both the Tevatron and the LHC data \cite{Buckley:2015lku}.  Individual limits are given for each
operator, by setting other operator coefficients to zero.  Marginalised
constraints are provided for $\Op{\varphi Q}^{\sss (3)}$, $\Op{tW}$, and
$\Op{tG}$, while the remaining operator constraints are too weak due to large
uncertainties in $pp\to t\bar tZ$ and $pp\to t\bar t\gamma$ measurements. One can see that $\Op{tG}$ is already significantly better constrained than its weak counterparts.
In addition, the $\Op{\varphi tb}$ operator gives rise to right handed $Wtb$
coupling, which is constrained at tree-level by top decay measurements and
indirectly at loop-level by $B$ meson decay and $h\to b\bar b$
\cite{Alioli:2017ces}. The electroweak and top-quark Yukawa operators $\Op{W}$,
$\Op{\varphi W}$, $\Op{t\varphi}$, $\Op{HW}$ and $\Op{HB}$ are constrained
by a combined fit including Higgs data and TGC measurements at both LEP and
LHC, presented in Ref.~\cite{Butter:2016cvz}.  For the Yukawa operator
$\mathcal{O}_{t\varphi}$, we follow the approach in
Ref.~\cite{Maltoni:2016yxb}, and update the analysis with the recent $t\bar{t}H$
measurements at 13 TeV in
Refs.~\cite{CMS:2017lgc,CMS:2017vru,Aaboud:2017jvq,Aaboud:2017rss}, obtaining a confidence interval of $\Cp{t\phi}\subset[-6.5,1.3]$. 
Note that we do not use the $gg\to H$ process.  Even though this process
could impose strong constraints on the coefficient of $\mathcal{O}_{t\varphi}$,
the effect is loop-induced, and so we consider it as an indirect constraint. The constraints on the color singlet and octet four-fermion operators are obtained from single-top and $t\bar{t}$ measurements~\cite{Zhang:2017mls} respectively. Although the color octet operator interferes with the SM $q\bar{q}\to t\bar{t}$ amplitude, the sensitivity of the process to this operator is diluted by the dominantly $gg$-induced SM component.
Even though this operator does not interfere with the SM single-top amplitude, the sensitivity from the pure EFT squared contribution is still significantly better than that of $t\bar{t}$. Combining a set of LHC measurements of single top (and anti-top) production~\cite{Chatrchyan:2012ep,Khachatryan:2014iya,Aad:2014fwa,Aad:2015upn,Khachatryan:2016ewo,Aaboud:2016ymp,CMS:2016ayb,Aaboud:2017pdi}, we obtain a significant improvement on the confidence interval, $\Cp{Qq}^{\sss(3,8)}\subset[-1.40,1.20]$. The cross-section dependence is obtained from our model implementation at NLO in QCD. Finally, the precision electroweak measurements provide indirect limits on top-quark operators at the one-loop level.  Electroweak operators to which they
mix under RG running are required to be included in a global fit, but constraints on top-quark operators can be obtained by marginalising over these operators \cite{Zhang:2012cd}.  
\begin{table}
\begin{centering}
\begin{center}
\begin{tabular}{|llllll|}
	\hline
	Op. & TF (I) & TF (M) & RHCC (I) tree/loop & SFitter (I) & PEWM\footnote{marginalized over $\OO_{\varphi WB}$ and $\OO_{\varphi D}$.}
	\\\hline
	$\Op{W}$  &    &   & &  [-0.18,0.18] & 
	\\
	$\Op{HW}$ &  &   & & [-0.32,1.62] &
    	\\
	$\Op{HB}$ &  &   & & [-2.11,1.57] &
   	 \\
	$\Op{\varphi W}$ & & &  & [-0.39,0.33] &
	\\
	\hline
	$\Op{\varphi tb}$ & &    &  [-5.28,5.28]/[-0.046,0.040]   & &
	\\
	$\Op{\varphi Q}^{\sss (3)}$ & [-2.59,1.50] & [-4.19,2.00]& &  &$-1.0\pm2.7$ \footnote{Assuming $C_{\varphi Q}^{(3)}=-C_{\varphi Q}^{(3)}$. Same for $\OO_{\varphi Q}^{(1)}$.}
	\\
	$\Op{\varphi Q}^{\sss (1)}$ & [-3.10,3.10] &&& &$1.0\pm2.7$
	\\
	$\Op{\varphi t}$ & [-9.78,8.18] &&& & $1.8\pm3.8$
	\\
	$\Op{tW}$ & [-2.49,2.49] & [-3.99,3.40]  && & $-0.4\pm2.4$
	\\
	$\Op{tB}$ & [-7.09,4.68] &&& & $4.8\pm10.6$
	\\
	$\Op{tG}$ & [-0.24,0.53] & [-1.07,0.99] && &
	\\
	$\Op{t\varphi}$ &  &  &  & [-18.2,6.30] &
	 \\
	\hline
	$\Op{Qq}^{\sss (3,1)}$ & [-0.40,0.60]  & [−0.66,1.24] & & &
    	\\
	$\Op{Qq}^{\sss (3,8)}$ & [-4.90,3.70]  & [−6.06,6.73] & & &
	\\\hline
\end{tabular}
\end{center}

\caption{Limits on operator coefficients, from
the TopFitter (TF) Collaboration (individual (I) and marginalised (M)) \cite{Buckley:2015lku} supplemented with constraints on $c_{\sss Qq}^{\sss(3,8)}$ from $t\bar{t}$ measurements quoted in Ref.~\cite{Zhang:2017mls},
right-handed charged currents (RHCC, individual) \cite{Alioli:2017ces},
the SFitter group (individual) \cite{Butter:2016cvz},
and precision electroweak measurements \cite{Zhang:2012cd}.
$\Lambda=1$ TeV is assumed. \label{tab:constraints}
}
\end{centering}
\end{table}
\section{Calculation setup and numerical results}
\label{sec:setup}
Our computation is performed within the {\sc MG5\_aMC} framework
\cite{Alwall:2014hca} with all the elements
entering the NLO computations available automatically starting from the
SMEFT Lagrangian \cite{Alloul:2013bka, deAquino:2011ub, Degrande:2011ua,Degrande:2014vpa,
Hirschi:2011pa,Frederix:2009yq}. In addition to the SM-like scale and PDF uncertainties, 
we also compute the uncertainties due to missing higher orders in the $\alpha_s$
expansion of the EFT operators, following the procedure described in \cite{Maltoni:2016yxb}. Therein, a second renormalisation scale, $\mu_{EFT}$, is introduced such that the EFT renormalisation scale can be varied independently from the QCD one.

The cross section can be parametrised as:
\begin{flalign}
	\sigma=\sigma_{SM}+\sum_i\frac{1{\rm TeV}^2}{\Lambda^2}C_i\sigma_i
	+\sum_{i\leq j}
	\frac{1{\rm TeV}^4}{\Lambda^4}C_iC_j\sigma_{ij}.
	\label{eq:xsecpara}
\end{flalign} We provide results for $\sigma_i$ and $\sigma_{ij}$ for the LHC at 13 TeV in the 5-flavour scheme. 
Results are obtained with NNPDF3.0 LO/NLO PDFs \cite{Ball:2014uwa}, for
LO and NLO results respectively; input parameters are
\begin{flalign}
	&m_t=172.5\ \mathrm{GeV}\,, \quad
	m_H=125\ \mathrm{GeV}\,, \quad
	m_Z=91.1876\ \mathrm{GeV}\,, \\
	&\alpha_{EW}^{-1}=127.9\,, \quad
	G_F=1.16637\times10^{-5}\ \mathrm{GeV}^{-2}\,.
	\label{eq:input}
\end{flalign} Central scales for $\mu_R,\mu_F,\mu_{EFT}$ are chosen as $(m_t+m_H)/4$ for the $tHj$ process following the discussion in \cite{Demartin:2015uha}, and correspondingly $(m_t+m_Z)/4$ for the $tZj$. Three types of uncertainties are computed. The first is the standard
scale uncertainty, obtained by independently setting
$\mu_R$ and $\mu_F$ to $\mu/2$, $\mu$ and $2\mu$, where $\mu$ is the central
scale, obtaining nine $(\mu_R,\mu_F)$ combinations. The second uncertainty comes
from the NNPDF3.0 sets. The third one is the EFT scale uncertainty,
representing the missing higher-order corrections to the operators, obtained by
varying $\mu_{EFT}$, taking into account the effect of running of the Wilson coefficients from the central scale up to this new scale. This uncertainty is obtained for contributions involving the $(\Op{t\phi},\Op{tW},\Op{tB})$ operators which are the ones that run under QCD as discussed in Section~\ref{section:Operators}. 

\subsection{Inclusive results
\label{sec:inclusive}}
Results for the $tHj$ and $tZj$ cross section from individual operators are shown in Tables \ref{tHj:sigma} and \ref{tZj:sigma} 
along with the corresponding uncertainties and $K$-factors. We note here that our results refer to the sum 
of the top and anti-top contributions. Central values for the cross-terms between the different operators are reported in Tables~\ref{tHj:sigma_crs} and ~\ref{tZj:sigma_crs}.
Several observations are in order. First we notice that the $K$-factors 
vary a lot between operator contributions. As we work in the 5-flavour scheme, the $b$-quark is massless, and therefore 
$\Op{\phi tb}$ does not interfere with the SM or any of the other operators. We also find that typically the relative
EFT contributions to $tHj$ are larger than for $tZj$, as the Higgs always couples to the top or the 
gauge bosons, whilst the $Z$ can be also be emitted from the light quark lines thus being unaffected by 
modifications of the top-$Z$ and triple gauge boson interactions. For $tZj$ some interferences between operators are suppressed and our results can suffer from rather large statistical errors as these contributions are extracted from Monte Carlo runs which involve all relevant SM, $\mathcal{O}\left(1/\Lambda^2 \right)$ and $\mathcal{O}\left(1/\Lambda^4 \right)$ terms arising from a given combination of couplings. 
\begin{table}[ht!]
\centering
    \renewcommand{\arraystretch}{1.4}
  {  \centering\small
    \begin{tabular}{l|lll}
        $\sigma$ [fb]& LO& NLO&K-factor\tabularnewline
        \hline\hline
        $\sigma_{SM}$&
        $57.56(4)^{+11.2\%}_{-7.4\%} \pm10.2\%$&
        $75.87(4)^{+2.2\%}_{-6.4\%} \pm1.2\%$&
        $1.32$
        \tabularnewline\hline
        $\sigma_{\sss \phi W}$&
        $8.12(2)^{+13.1\%}_{-9.3\%} \pm9.3\%$&
        $7.76(2)^{+7.0\%}_{-6.3\%} \pm1.0\%$&
        $0.96$
        \tabularnewline
        $\sigma_{\sss \phi W,\phi W}$&
        $5.212(7)^{+10.6\%}_{-6.8\%} \pm10.2\%$&
        $6.263(7)^{+2.6\%}_{-7.8\%} \pm1.3\%$&
        $1.20$
        \tabularnewline\hline
        $\sigma_{\sss t\phi}$&
        $-1.203(6)^{+12.0\%}_{-15.6\%} \pm8.9\%$&
        $-0.246(6)^{+144.5[31.4]\%}_{-157.8[19.0]\%} \pm2.1\%$&
        $0.20$
        \tabularnewline
        $\sigma_{\sss t\phi,t\phi}$&
        $0.6682(9)^{+12.7\%}_{-8.9\%} \pm9.6\%$&
        $0.7306(8)^{+4.6[0.6]\%}_{-7.3[0.2]\%} \pm1.0\%$&
        $1.09$
        \tabularnewline\hline
        $\sigma_{\sss tW}$&
        $19.38(6)^{+13.0\%}_{-9.3\%} \pm9.4\%$&
        $22.18(6)^{+3.8[0.4]\%}_{-6.8[0.9]\%} \pm1.0\%$&
        $1.14$
        \tabularnewline
        $\sigma_{\sss tW,tW}$&
        $46.40(8)^{+9.3\%}_{-5.5\%} \pm11.1\%$&
        $71.24(8)^{+7.4[1.5]\%}_{-14.0[6.9]\%} \pm1.9\%$&
        $1.54$
        \tabularnewline\hline
        $\sigma_{\sss \phi Q^{(3)}}$&
        $-3.03(3)^{+0.0\%}_{-2.2\%} \pm15.4\%$&
        $-10.04(4)^{+11.1\%}_{-8.9\%} \pm1.8\%$&
        $3.31$
        \tabularnewline
        $\sigma_{\sss \phi Q^{(3)},\phi Q^{(3)}}$&
        $11.23(2)^{+9.4\%}_{-5.6\%} \pm11.2\%$&
        $15.28(2)^{+5.0\%}_{-10.9\%} \pm1.8\%$&
        $1.36$
        \tabularnewline\hline
        $\sigma_{\sss \phi tb}$&
        $0$&
        $0$&
        $-$
        \tabularnewline
        $\sigma_{\sss \phi tb,\phi tb}$&
        $2.752(4)^{+9.4\%}_{-5.5\%} \pm11.3\%$&
        $3.768(4)^{+5.0\%}_{-10.9\%} \pm1.8\%$&
        $1.54$
        \tabularnewline\hline
        $\sigma_{\sss HW}$&
        $-3.526(4)^{+5.6\%}_{-9.5\%} \pm10.9\%$&
        $-5.27(1)^{+6.5\%}_{-2.9\%} \pm1.5\%$&
        $1.50$
        \tabularnewline
        $\sigma_{\sss HW,HW}$&
        $0.9356(4)^{+7.9\%}_{-4.0\%} \pm12.3\%$&
        $1.058(1)^{+4.8\%}_{-11.9\%} \pm2.3\%$&
        $1.13$
        \tabularnewline\hline
        $\sigma_{\sss tG}$&
         \multicolumn{2}{c}{$-0.418(5)^{+12.3\%}_{-9.8\%} \pm1.1\%$}&
        $-$
        \tabularnewline
        $\sigma_{\sss tG,tG}$&
         \multicolumn{2}{c}{$1.413(1)^{+21.3\%}_{-30.6\%} \pm2.5\%$}&        
        $-$
        \tabularnewline\hline
        $\sigma_{\sss Qq^{(3,1)}}$&
        $-22.50(5)^{+8.0\%}_{-11.8\%} \pm9.7\%$&
        $-20.10(5)^{+13.8\%}_{-13.3\%} \pm1.1\%$&
        $0.89$
        \tabularnewline
        $\sigma_{\sss Qq^{(3,1)},Qq^{(3,1)}}$&
        $69.78(3)^{+8.0\%}_{-4.1\%} \pm12.1\%$&
        $62.20(3)^{+11.5\%}_{-15.9\%} \pm2.3\%$&
        $0.89$
        \tabularnewline\hline
        $\sigma_{\sss Qq^{(3,8)}}$&
        $-$&
        $0.25(3)^{+25.4\%}_{-27.1\%} \pm4.7\%$&
        $-$
        \tabularnewline
        $\sigma_{\sss Qq^{(3,8)},Qq^{(3,8)}}$&
        $15.53(2)^{+8.0\%}_{-4.1\%} \pm12.1\%$&
        $14.07(2)^{+11.0\%}_{-15.7\%} \pm2.1\%$&
        $0.91$
        \tabularnewline\hline
    \end{tabular}}
  \caption{Cross-section results for $tHj$ at 13 TeV, following the parametrisation of Eq.~( \ref{eq:xsecpara}). Central values are quoted followed by the upper and lower scale uncertainty bands obtained by varying the renormalisation scale between half and twice the central value,  the EFT scale uncertainty where relevant and finally the PDF uncertainty. The MC error on the last digit is shown in the bracket.}
        \label{tHj:sigma}
\end{table}

\begin{table}[ht!]
\centering
  \renewcommand{\arraystretch}{1.4}
    \small
    \begin{tabular}{l|lll}
        $\sigma$ [fb]& LO& NLO&K-factor\tabularnewline
        \hline
        \hline
        $\sigma_{SM}$&
        $660.8(4)^{+13.7\%}_{-9.6\%} \pm9.7\%$&
        $839.1(5)^{+1.1\%}_{-5.1\%} \pm1.0\%$&
        1.27
        \tabularnewline\hline
        $\sigma_{\sss  W}$&
        $-7.87(7)^{+8.4\%}_{-12.6\%} \pm9.7\%$&
        $-8.77(8)^{+8.5\%}_{-4.3\%} \pm1.1\%$&
        1.12
        \tabularnewline
        $\sigma_{\sss W, W}$& 
        $34.58(3)^{+8.2\%}_{-3.9\%} \pm13.0\%$&
        $43.80(4)^{+6.6\%}_{-15.1\%} \pm2.8\%$&
        1.27
        \tabularnewline\hline
        $\sigma_{\sss tB}$&
        $2.23(2)^{+14.7[0.9]\%}_{-10.7[1.0]\%} \pm9.4\%$ &
        $2.94(2)^{+2.3[0.4]\%}_{-3.0[0.7]\%} \pm1.1\%$&
        1.32        
        \tabularnewline
        $\sigma_{\sss tB,tB}$&
        $2.833(2)^{+10.5[1.7]\%}_{-6.3[1.9]\%} \pm11.1\%$&
        $4.155(3)^{+4.7[0.9]\%}_{-10.1[1.4]\%} \pm1.7\%$&
        1.47
        \tabularnewline\hline
        $\sigma_{\sss tW}$&
        $2.66(4)^{+18.8[0.9]\%}_{-15.3[1.0]\%} \pm11.4\%$&
        $13.0(1)^{+15.8[2.1]\%}_{-22.8[0.0]\%} \pm1.2\%$&
        4.90
        \tabularnewline
        $\sigma_{\sss tW,tW}$&
        $48.16(4)^{+10.0[1.7]\%}_{-5.8[1.9]\%} \pm11.3\%$&
        $80.00(4)^{+7.9[1.3]\%}_{-14.7[1.6]\%} \pm1.9\%$&
        1.66
        \tabularnewline\hline
        $\sigma_{\sss \phi dtR}$&
        $4.20(1)^{+14.9\%}_{-10.9\%} \pm9.3\%$&
        $4.94(2)^{+3.4\%}_{-6.7\%} \pm1.0\%$&
        1.18
        \tabularnewline
        $\sigma_{\sss \phi dtR,\phi dtR}$&
        $0.3326(3)^{+13.6\%}_{-9.5\%} \pm9.6\%$&
        $0.4402(5)^{+3.7\%}_{-9.3\%} \pm1.0\%$&
        1.32
        \tabularnewline\hline
        $\sigma_{\sss \phi Q}$&
        $14.98(2)^{+14.5\%}_{-10.5\%} \pm9.4\%$&
        $18.07(3)^{+2.3\%}_{-1.6\%} \pm1.0\%$&
        1.21
        \tabularnewline
        $\sigma_{\sss \phi Q,\phi Q}$&
        $0.7442(7)^{+14.1\%}_{-10.0\%} \pm9.5\%$&
        $1.028(1)^{+2.8\%}_{-7.3\%} \pm1.0\%$&
        1.38
        \tabularnewline\hline
        $\sigma_{\sss \phi Q^{(3)}}$&
        $130.04(8)^{+13.8\%}_{-9.8\%} \pm9.5\%$&
        $161.4(1)^{+0.9\%}_{-4.8\%} \pm1.0\%$ &
        1.24
        \tabularnewline
        $\sigma_{\sss \phi Q^{(3)},\phi Q^{(3)}}$&
        $17.82(2)^{+11.7\%}_{-7.5\%} \pm10.5\%$&
        $23.98(2)^{+3.7\%}_{-9.3\%} \pm1.4\%$&
        1.35
        \tabularnewline\hline
        $\sigma_{\sss \phi tb}$&
        $0$&
        $0$&
        $-$
        \tabularnewline
        $\sigma_{\sss \phi tb,\phi tb}$&
        $2.949(2)^{+10.5\%}_{-6.2\%} \pm11.1\%$&
        $4.154(4)^{+5.1\%}_{-11.2\%} \pm1.8\%$&
        1.41
        \tabularnewline\hline
        $\sigma_{\sss HW}$&
        $-5.16(6)^{+7.8\%}_{-12.0\%} \pm10.5\%$&
        $-6.88(8)^{+6.4\%}_{-2.0\%} \pm1.4\%$&
        1.33
        \tabularnewline
        $\sigma_{\sss HW,HW}$&
        $0.912(2)^{+9.4\%}_{-5.2\%} \pm12.0\%$&
        $1.048(2)^{+5.2\%}_{-12.8\%} \pm2.1\%$&
        1.15
        \tabularnewline\hline
        $\sigma_{\sss HB}$&
        $-3.015(9)^{+9.9\%}_{-13.9\%} \pm9.5\%$&
        $-3.76(1)^{+5.2\%}_{-1.0\%} \pm1.0\%$&
        1.25
        \tabularnewline
        $\sigma_{\sss HB,HB}$&
        $0.02324(6)^{+12.7\%}_{-8.5\%} \pm9.9\%$&
        $0.02893(6)^{+2.3\%}_{-7.5\%} \pm1.1\%$&
        1.24
        \tabularnewline\hline
        $\sigma_{\sss  tG}$&
        \multicolumn{2}{c}{$0.45(2)^{+93.0\%}_{-148.8\%} \pm4.9\%$}
        &
        $-$
        \tabularnewline
        $\sigma_{\sss tG,tG}$&
        \multicolumn{2}{c}{$2.251(4)^{+20.9\%}_{-30.0\%} \pm2.5\%$}
        &
        $-$
        \tabularnewline\hline
         
        $\sigma_{\sss Qq^{(3,1)}}$&
        $-393.5(5)^{+8.1\%}_{-12.3\%} \pm10.0\%$  &
        $-498(1)^{+8.9\%}_{-3.2\%} \pm1.2\%$&
        $1.26$
        \tabularnewline
        
        $\sigma_{\sss Qq^{(3,1)},Qq^{(3,1)}}$&
        $462.25(3)^{+8.4\%}_{-4.1\%} \pm12.7\%$  &
        $545.50(5)^{+7.4\%}_{-17.4\%} \pm2.9\%$    &
        $1.18$
        \tabularnewline\hline
        
        $\sigma_{\sss Qq^{(3,8)}}$&
        $0$&
        $-0.9(3)^{+23.3\%}_{-26.3\%} \pm19.2\%$&
        $-$
        \tabularnewline
        
        $\sigma_{\sss Qq^{(3,8)},Qq^{(3,8)}}$&
        $102.73(5)^{+8.4\%}_{-4.1\%} \pm12.7\%$&
        $111.18(5)^{+9.3\%}_{-18.4\%} \pm2.8\%$&
        $1.08$
        \tabularnewline\hline
    \end{tabular}

        \caption{Cross-section results for $tZj$ at 13 TeV, following the parametrisation of Eq.~ (\ref{eq:xsecpara}). Central values are quoted followed by the upper and lower scale uncertainty bands obtained by varying the renormalisation scale between half and twice the central value, the EFT scale uncertainty where relevant and finally the PDF uncertainty.  The MC error on the last digit is shown in the bracket. }
        \label{tZj:sigma}
\end{table}

\begin{figure}[h]
    \centering
    \includegraphics[width=0.45\linewidth]{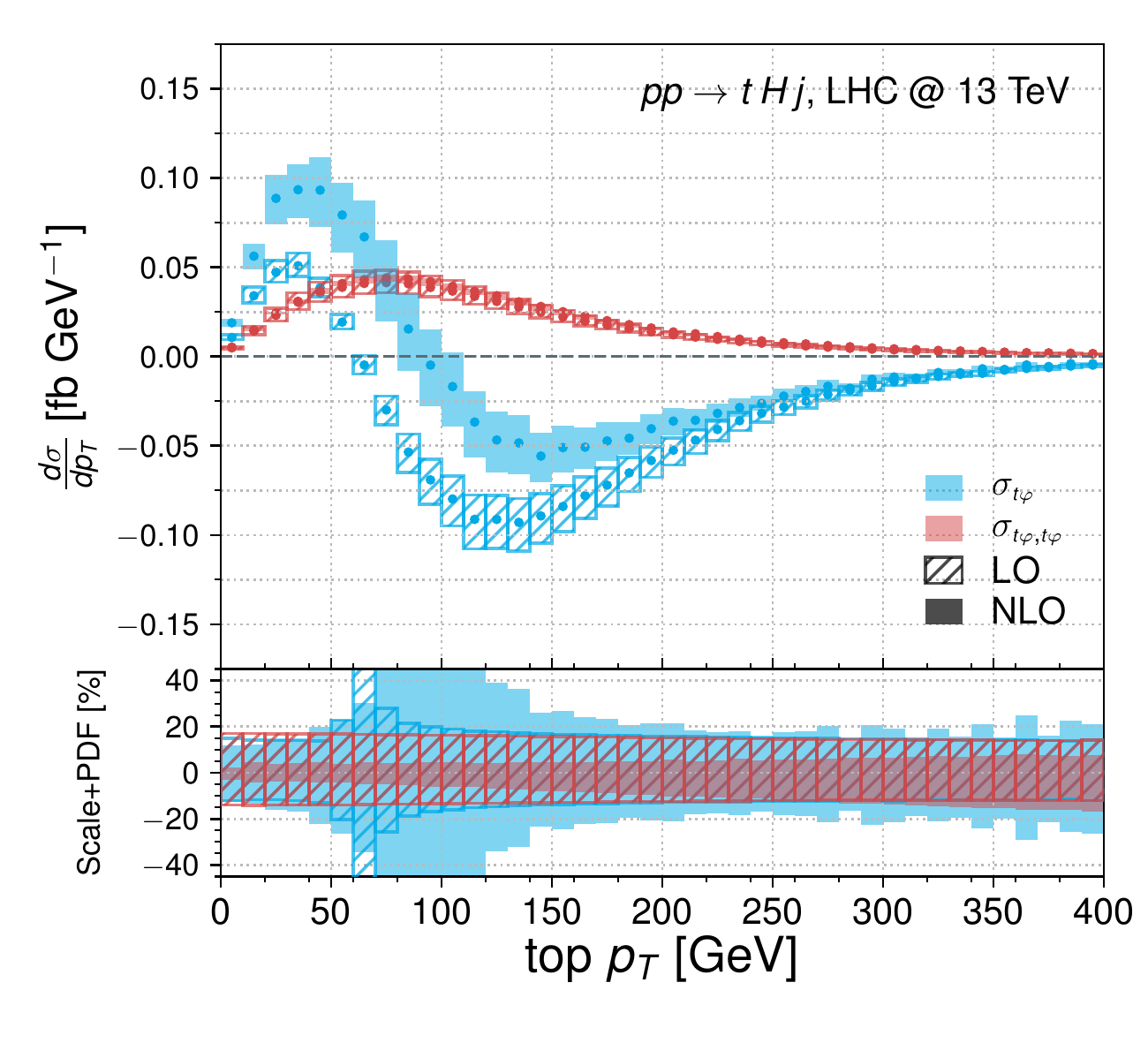}
    \includegraphics[width=0.45\linewidth]{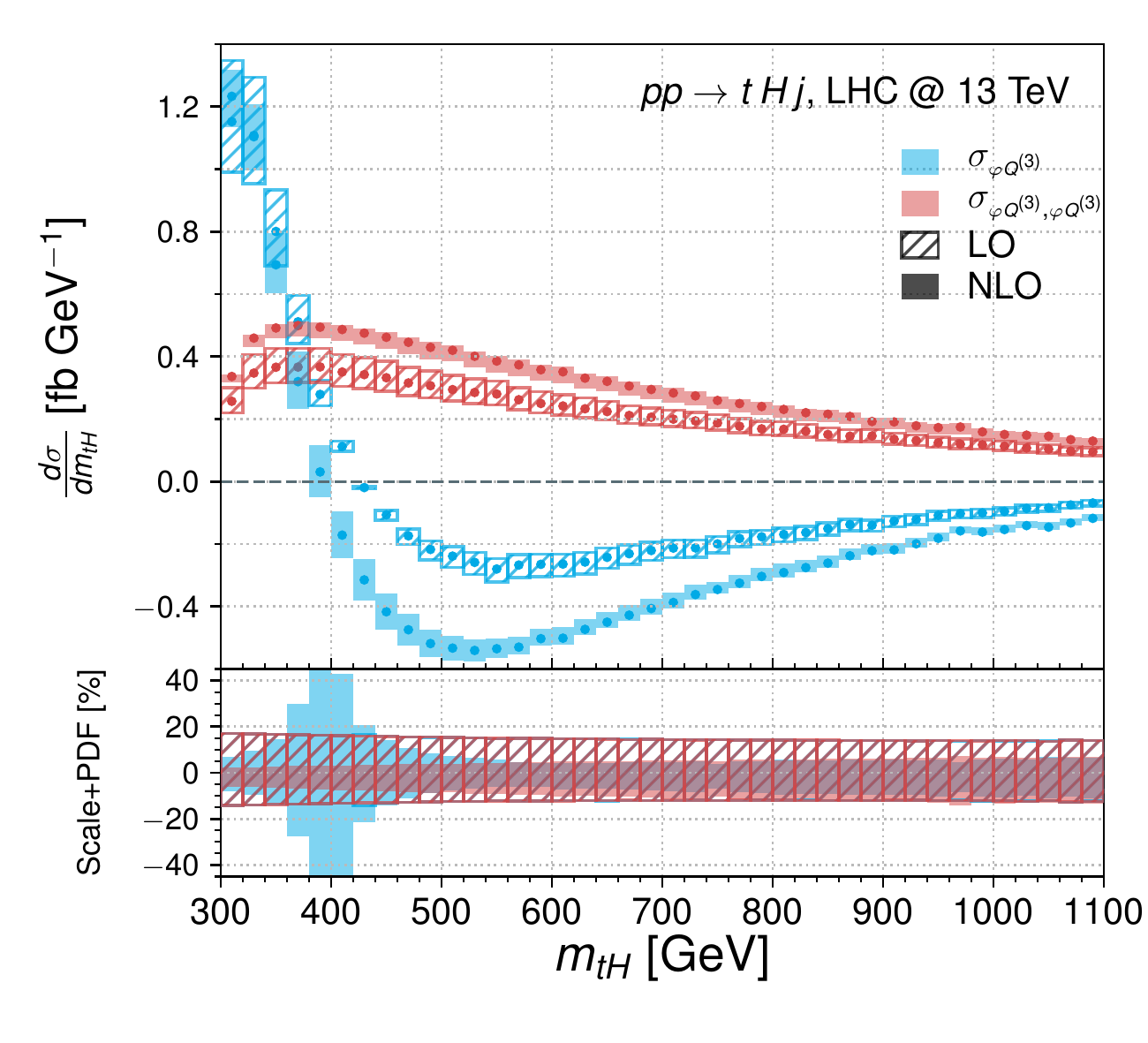}
    \includegraphics[width=0.45\linewidth]{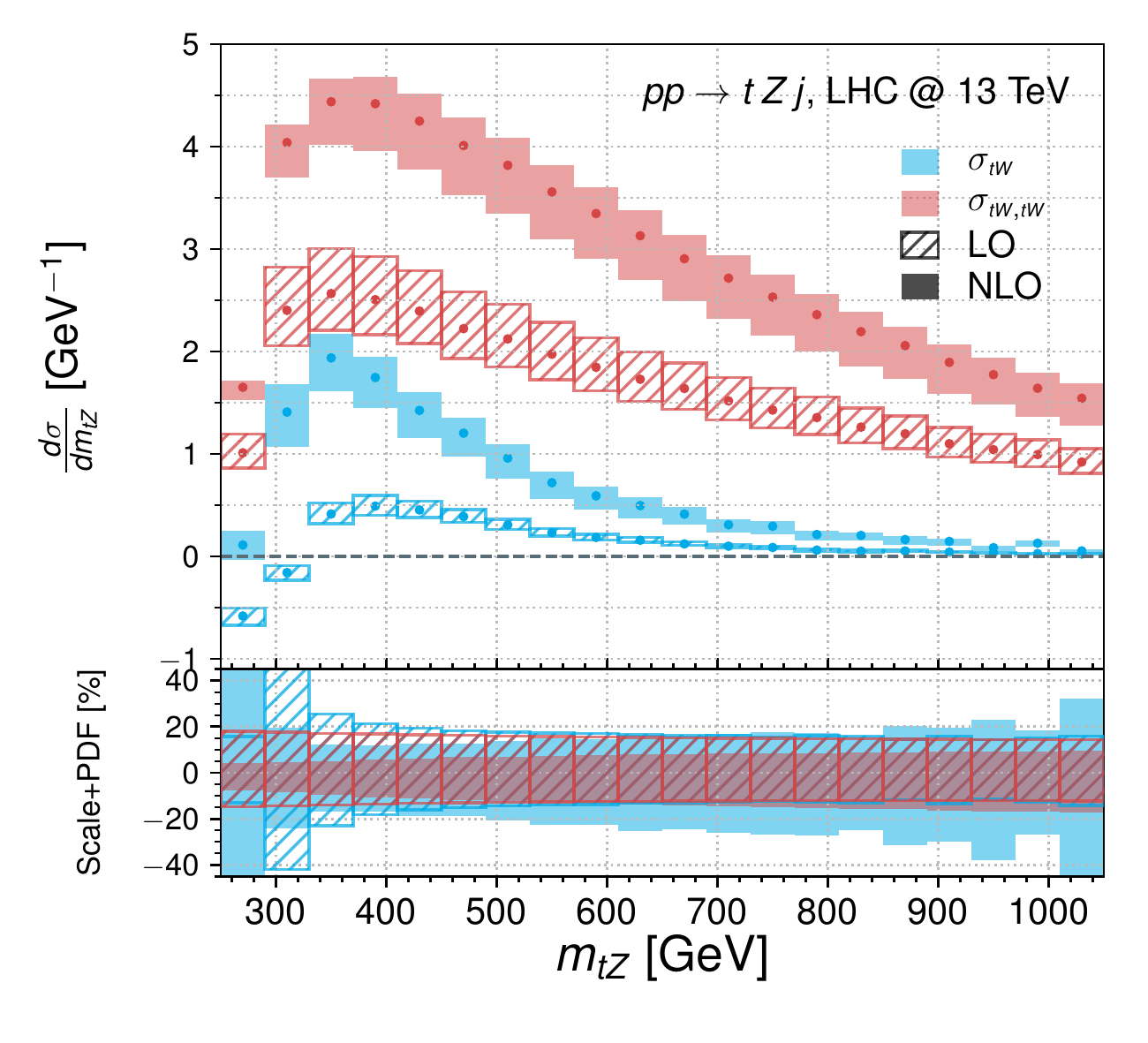}
    \caption{\label{fig:top_pt_sqint}
    Differential cross-section contributions to $tHj$ from 
    $\Op{t\phi}$ and $\Op{\phi Q}^{\sss (3)}$ and similarly for the $\Op{tW}$ 
    contribution to $tZj$, all for values of 1 TeV$^{-2}$ of the corresponding Wilson coefficient. Hatched and solid bars represent the LO and NLO 
    predictions respectively. The subplots show the relative theory 
    uncertainty from scale variation and PDFs of each contribution. 
    }
\end{figure}

In general, we see that the NLO corrections reduce the theory uncertainties and that the EFT scale uncertainty is typically subdominant. One striking case stands out in which the scale uncertainty for the inclusive interference contribution from $\Op{t\phi}$ to $tHj$ grows significantly. This can be understood by looking at the differential level and noticing that there is a very strong cancellation over the phase space such that the contribution to the total rate coming from the interference almost cancels. Figure~\ref{fig:top_pt_sqint} shows the top $p_T$ distributions of the interference and squared contributions at LO and NLO. Clearly, the cancellation is even more exact at NLO and leads to large scale uncertainties in the inclusive result and the unusual $K$-factor of 0.2. A partial cancellation effect is also present for the $\Op{\phi Q}^{\sss (3)}$ interference contribution at LO, which is reduced at NLO, leading to the correspondingly large $K$-factor. This is best seen from the top-Higgs invariant mass distribution also shown in Figure~\ref{fig:top_pt_sqint}.
As for $tZj$, we observe qualitatively similar results moving from LO to NLO. In some cases, for numerically very small contributions coming from interference terms between operators, the theory uncertainties are inflated due to lack of MC stats. The main unexpected result is the $K$-factor of 5 for the $\Op{tW}$ interference contribution. The top-$Z$ invariant mass distribution in Figure~\ref{fig:top_pt_sqint}, does indicate a cancellation over the full phase space which disappears at NLO. This is in part due to cancellations in the interference contributions to $tZj$ and $\bar{t}Zj$, which are summed over in our results. 
                
Considering the existing limits on the Wilson coefficients summarised in Table~\ref{tab:constraints} in combination with the information in Tables~\ref{tHj:sigma} and~\ref{tZj:sigma} suggests that there is still much room for observable deviations in both processes and therefore that they may be used to further constrain the SMEFT parameter space. For example, saturating the current limits on the weak dipole operators, $\Op{tW}$ and $\Op{tB}$, leads to 20\% deviations in the inclusive $tZj$ cross section at NLO while for $tHj$, the corresponding effects of $\Op{tW}$ and the top-Yukawa operator, $\Op{t\phi}$, are around 300\%. Deviations to $tZj$ are generally possible within current limits at the level of up to 20\% while, for $tHj$, order one effects can additionally be accommodated for the right handed charged current operator. Given the weak limits on the operators in question the large cross-section contributions are dominated by the EFT-squared term.

It is instructive to put these calculations into context by comparing to the $t$-channel single top production process, which is a common sub-process of both processes studied in this work. Table~\ref{tab:sensitivity} compares the interferences and squared contributions at NLO, relative to the SM, of the operators common to the $tHj$, $tZj$ and $t$-channel single top processes. We observe the expected enhancement of the relative contribution of the four fermion operators with respect to single top due to the higher kinematic thresholds involved. This is confirmed by adding a minimum $p_T^{t}$ such that the cross sections of $tj$ ($tZj$) becomes comparable to that of $tZj$ ($tHj$), which shows that $tj$ is likely to provide tighter constraints for these operators once the high $p_T^{t}$ regime is measured at 13 TeV. The behaviour of the sensitivity between the inclusive and high energy regions of each operator in $tZj$ is in line with the expectations from the $2\to2$ sub-amplitudes shown in Table~\ref{tab:bwtz}. One can confirm that the left handed quark current operator $\Op{\phi Q}^{\sss(3)}$ is significantly enhanced both at interference and squared level by the $p_T^t$ cut. Interestingly, this is in contrast to the case of single top production or top decay, where this operator only shifts the SM $Wtb$ vertex, not leading to any energy growth. The $tZj$ process provides a new source of energy dependence and therefore sensitivity to this operator that, as we will show in Section~\ref{sec:sensitivity}, may lead to potentially improved constraints in the near future.  In the case of the weak dipole and RHCC operators, $\Op{tW}$ and $\Op{\phi tb}$, the leading energy growth is confirmed to arise from the squared contribution. For $\Op{\phi tb}$, the high energy behaviour is enhanced with respect to single top. As discussed in Section~\ref{sec:energybehaviour}, the interferences of the configurations that have energy growth from $\Op{tW}$, are counterbalanced by an inverse dependence in the corresponding SM amplitudes, leading to the expected result that these pieces do not grow with energy. 

  \begin{table}[t]
      \centering
        \renewcommand{\arraystretch}{1.4}
    \small
    \begin{tabular}{l | c c c c c }
        & $tj$
        & $tj$
        & $tZj$
        & $tZj$
        & $tHj$\tabularnewline
        & 
        & ($p_T^{t}>350$ GeV) 
        & 
        & ($p_T^{t}>250$ GeV) 
        & \tabularnewline
        \hline
        \hline
        $\sigma_{SM}$
        & 224 pb 
        & 880 fb
        & 839 fb
        & 69 fb
        & 75.9 fb
        
        \tabularnewline\hline
        $r_{\sss tW}$
        & 0.028
        & 0.024
        & 0.0156
        & 0.0104
        & 0.292
        
        \tabularnewline
        $r_{\sss tW,tW}$
        & 0.016
        & 0.356
        & 0.096
        & 0.672
        & 0.940

        \tabularnewline\hline
        $r_{\sss \phi Q^{(3)}}$
        & 0.120
        & 0.120
        & 0.192
        & 0.686
        & -0.132
        
        \tabularnewline
        $r_{\sss \phi Q^{(3)},\phi Q^{(3)}}$
        & 0.0037
        & 0.0037
        & 0.023
        & 0.28
        & 0.21
        \tabularnewline
                \hline

        $r_{\sss \phi tb,\phi tb}$
        & 0.00090
        & 0.0008
        & 0.0050
        & 0.027
        & 0.050

        \tabularnewline\hline
        $r_{\sss  tG}$
        & 0.0003  
        & -0.01
        & 0.00053
        & -0.0048
        & -0.0055
        
        \tabularnewline
        $r_{\sss  tG,tG}$
        & 0.00062
        & 0.045
        & 0.0027
        & 0.022
        & 0.025

        \tabularnewline\hline
        $r_{\sss Qq^{(3,1)}}$
        & -0.353
        & -4.4
        & -0.595
        & -2.22
        & -0.39
      
        \tabularnewline
        $r_{\sss Qq^{(3,1)},Qq^{(3,1)}}$
        & 0.126
        & 11.5
        & 0.70
        & 5.08
        & 1.21
        
        \tabularnewline\hline
        
        
        $r_{\sss Qq^{(3,8)},Qq^{(3,8)}}$
        & 0.0308
        & 2.73
        & 0.16
        & 1.01
        & 1.08
\end{tabular}

      \caption{\label{tab:sensitivity} Comparison among the NLO sensitivities of $tj$ (inclusive and with $p_T^{t}>350$ GeV), $tZj$ (inclusive and with $p_T^{t}>250$ GeV), and $tHj$  to the six operators which are common to the three processes, {\it i.e.}, those entering in $tj$.  The interference term $r_i=\sigma_i/\sigma_{SM}$ (when non-zero) and the square $r_{i,i}=\sigma_{i,i}/\sigma_{SM}$ are given for each operator. $\sigma_i$ and $\sigma_{i,i}$ are defined in Eq.~(\ref{eq:xsecpara}).    }
  \end{table} 

\subsection{Differential distributions
\label{sec:differential}}
Given the promising effects observed in the inclusive cross-section predictions as well as Table~\ref{tab:sensitivity}, one expects even more striking deviations at differential level. This allows us to further investigate the energy dependence of the contributions from the various operators, comparing this to the expectations from the $2\to2$ helicity sub-amplitude calculations summarised in Tables~\ref{tab:bwth} and~\ref{tab:bwtz}. In order to showcase this, we present differential results in top $p_T$ and top-Higgs/$Z$ invariant mass for a number of benchmark scenarios, switching on one operator at a time to a value roughly saturating the tree-level, individual limits presented in Table~\ref{tab:constraints}. Individual limits are chosen for a fair representation since we are only switching on one operator at a time while indirect, loop-level limits are not taken into account since we are quantifying direct effects from SMEFT operators to these LHC processes. 

\begin{figure}[h]
    \centering
    \includegraphics[width=0.45\linewidth]{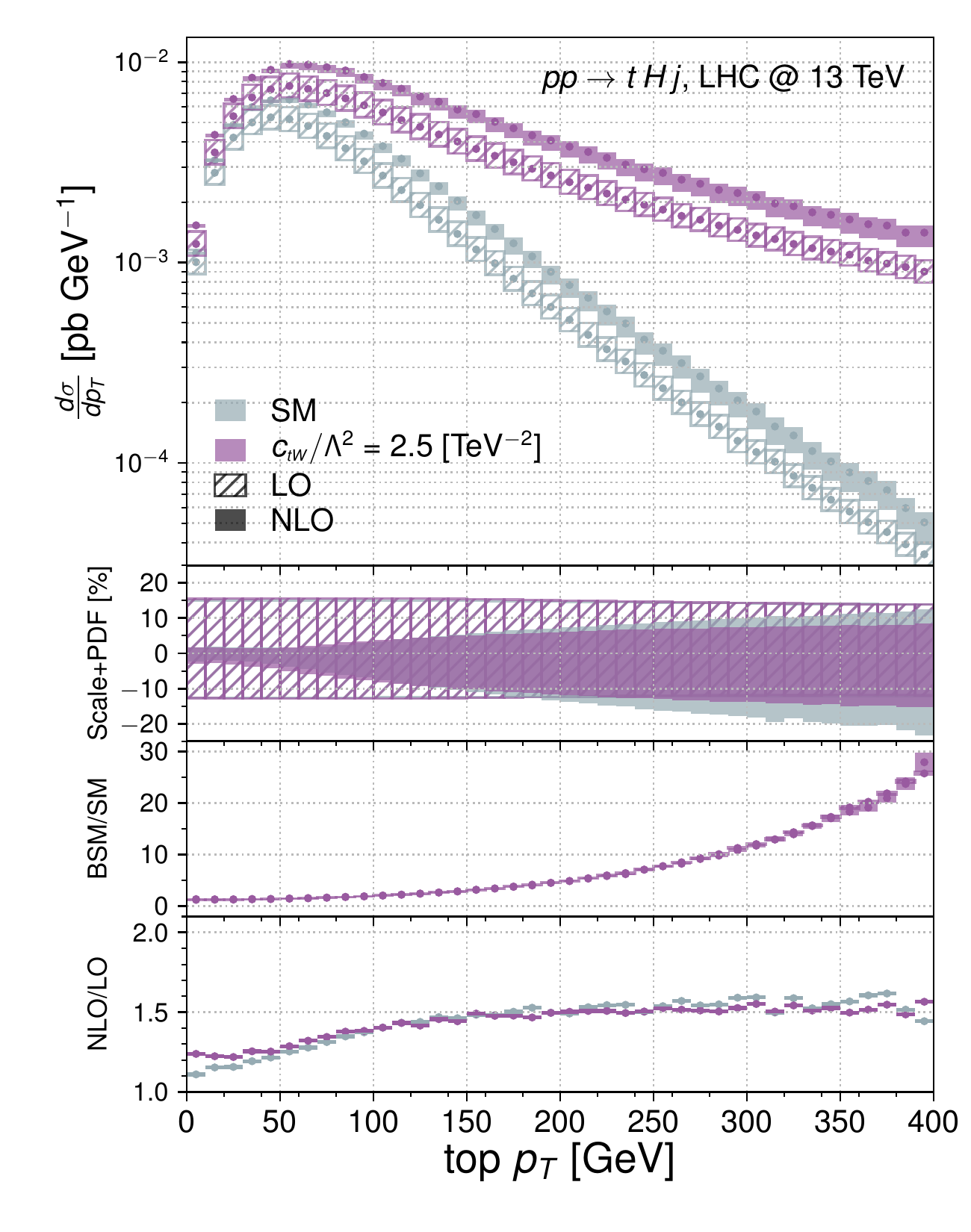}
    \includegraphics[width=0.45\linewidth]{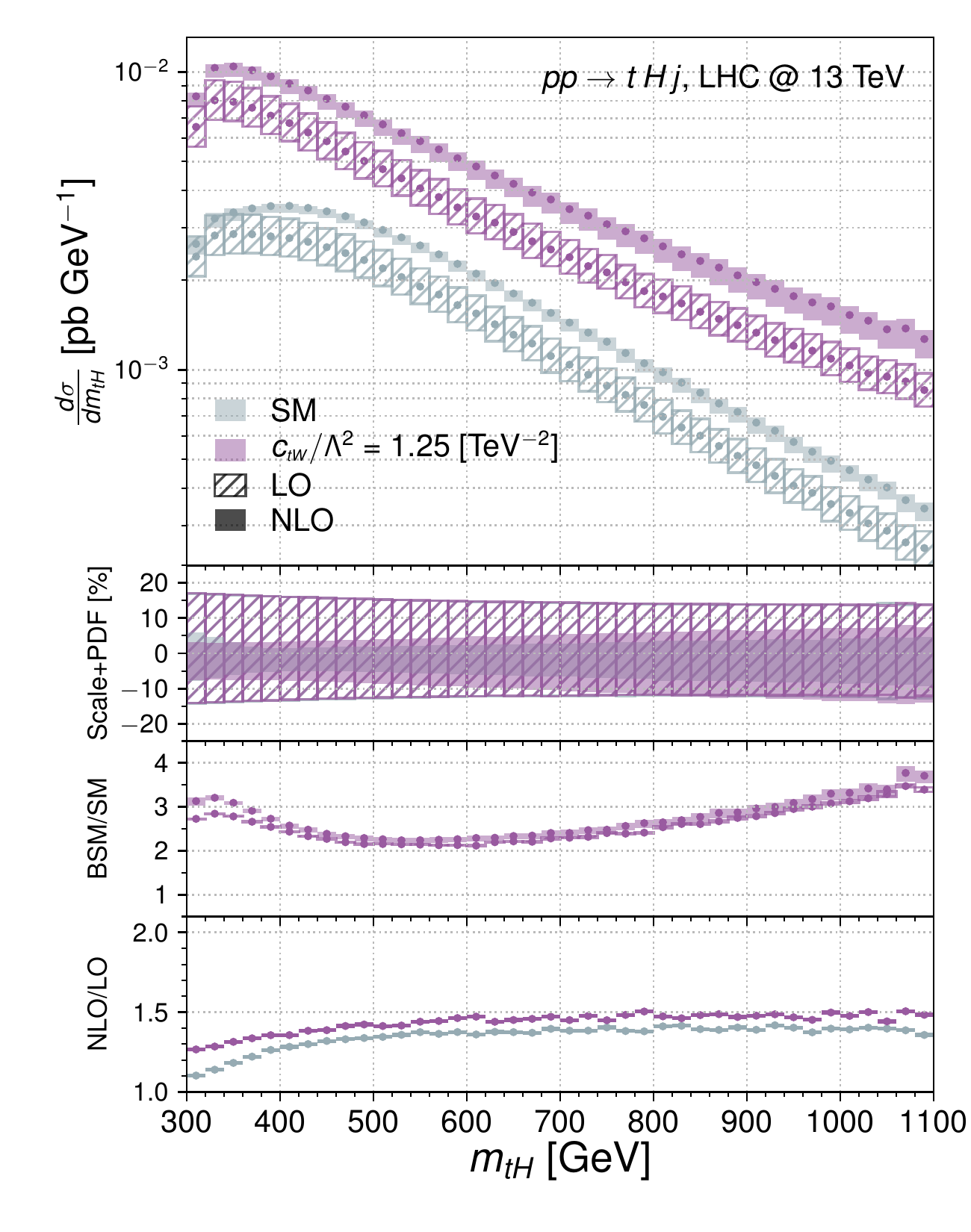}\\
    \includegraphics[width=0.45\linewidth]{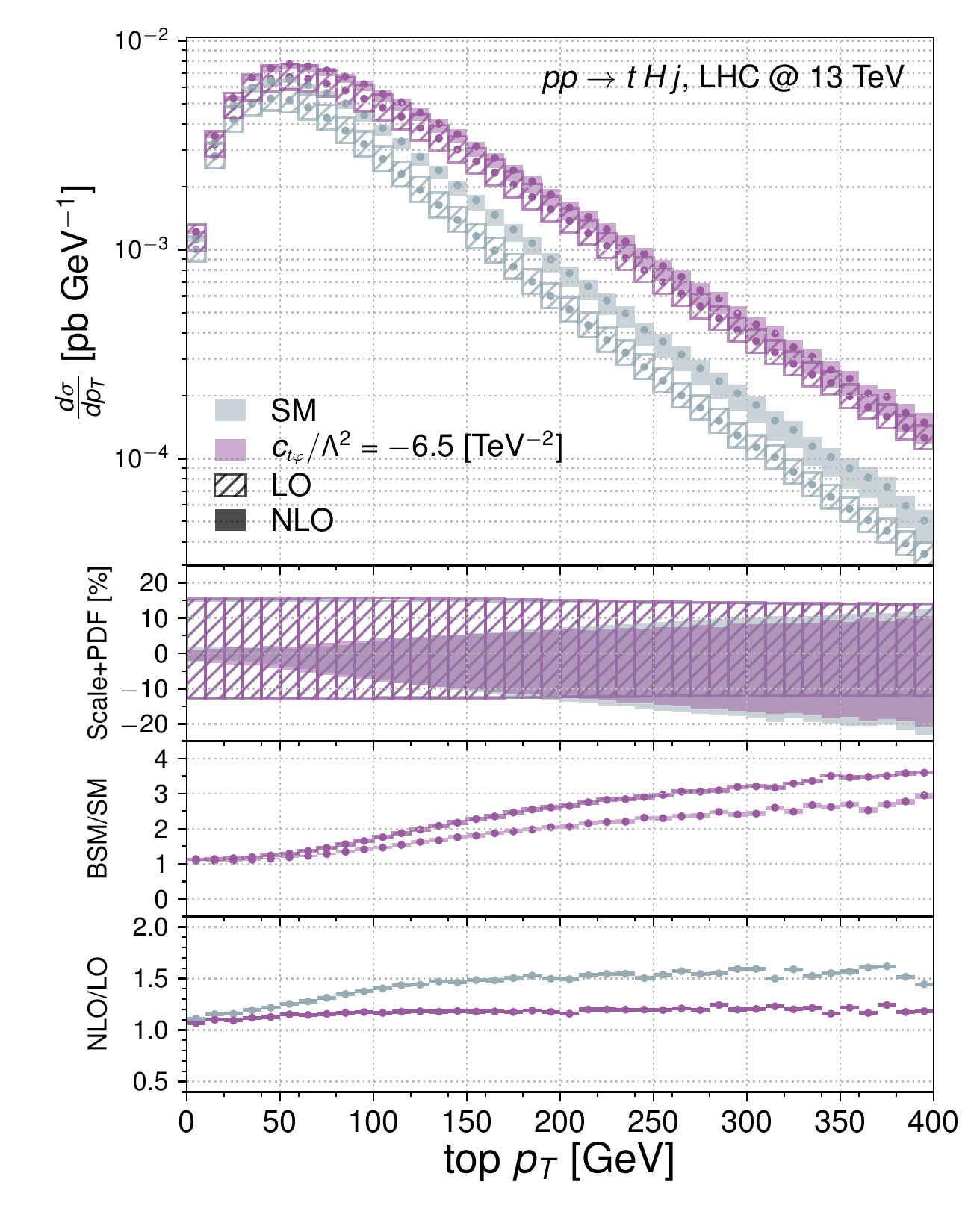}
    \includegraphics[width=0.45\linewidth]{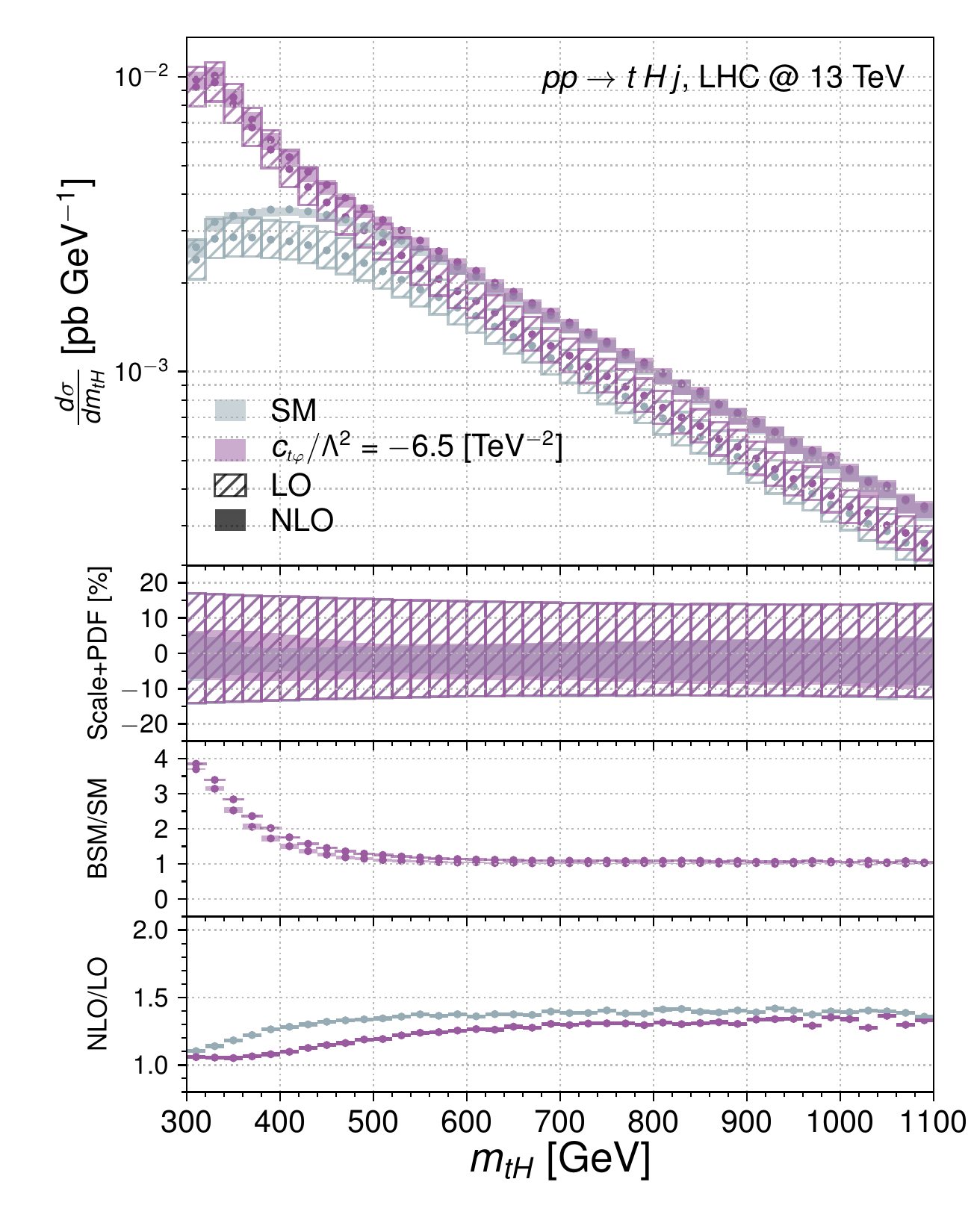}
    \caption{\label{fig:thj_distributions}
    Differential distributions of the top $p_T$ and top-Higgs system invariant mass 
    for the $tHj$ process for given values of the $\Op{tW}$  and $\Op{t\phi}$ operator coefficients roughly saturating current individual, direct limits. The lower insets 
    show the scale and PDF uncertainty bands, the ratio over the SM prediction and finally 
    the corresponding $K$-factor.  }
\end{figure}

\begin{figure}[h]
    \centering
    \includegraphics[width=0.45\linewidth]{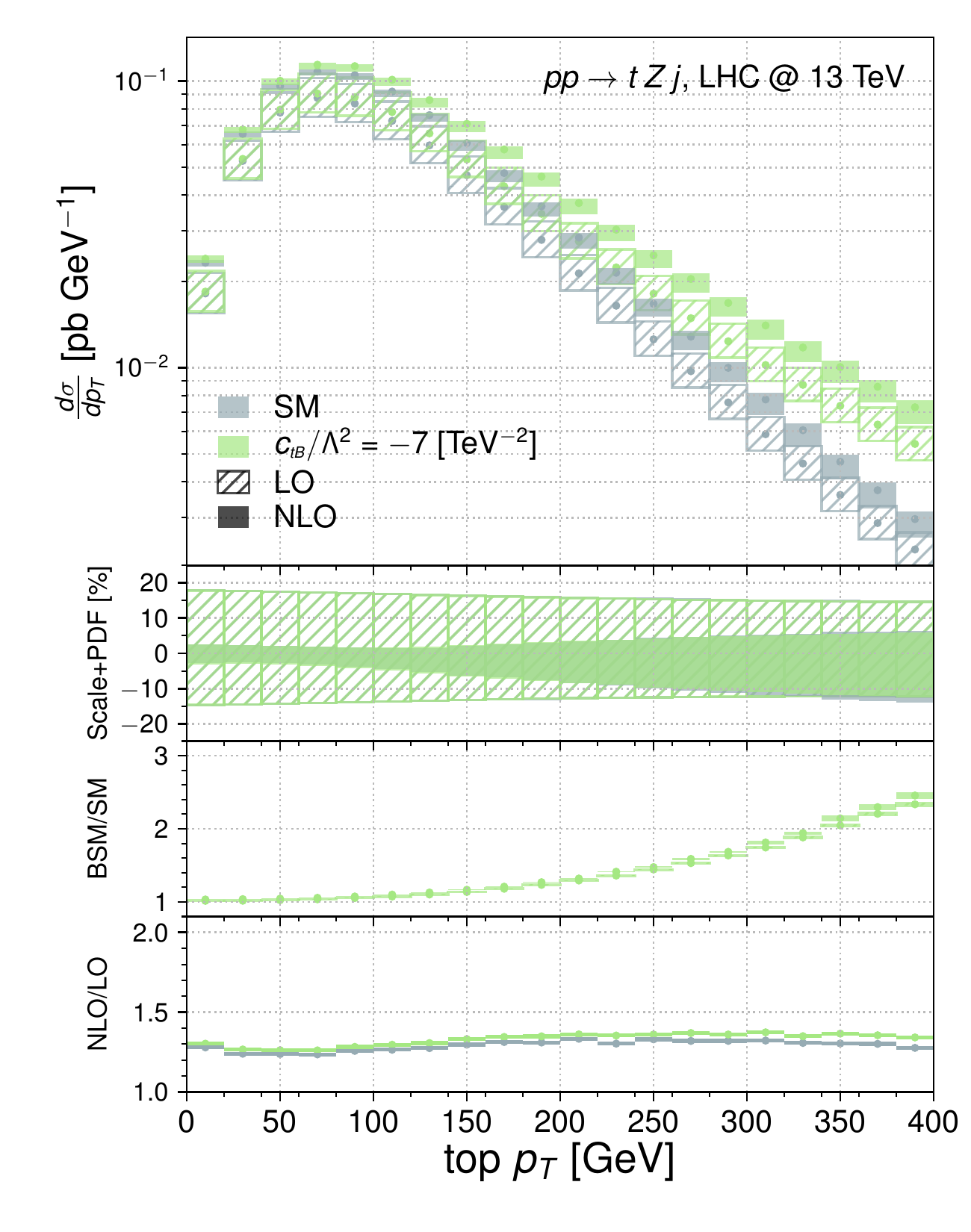}
    \includegraphics[width=0.45\linewidth]{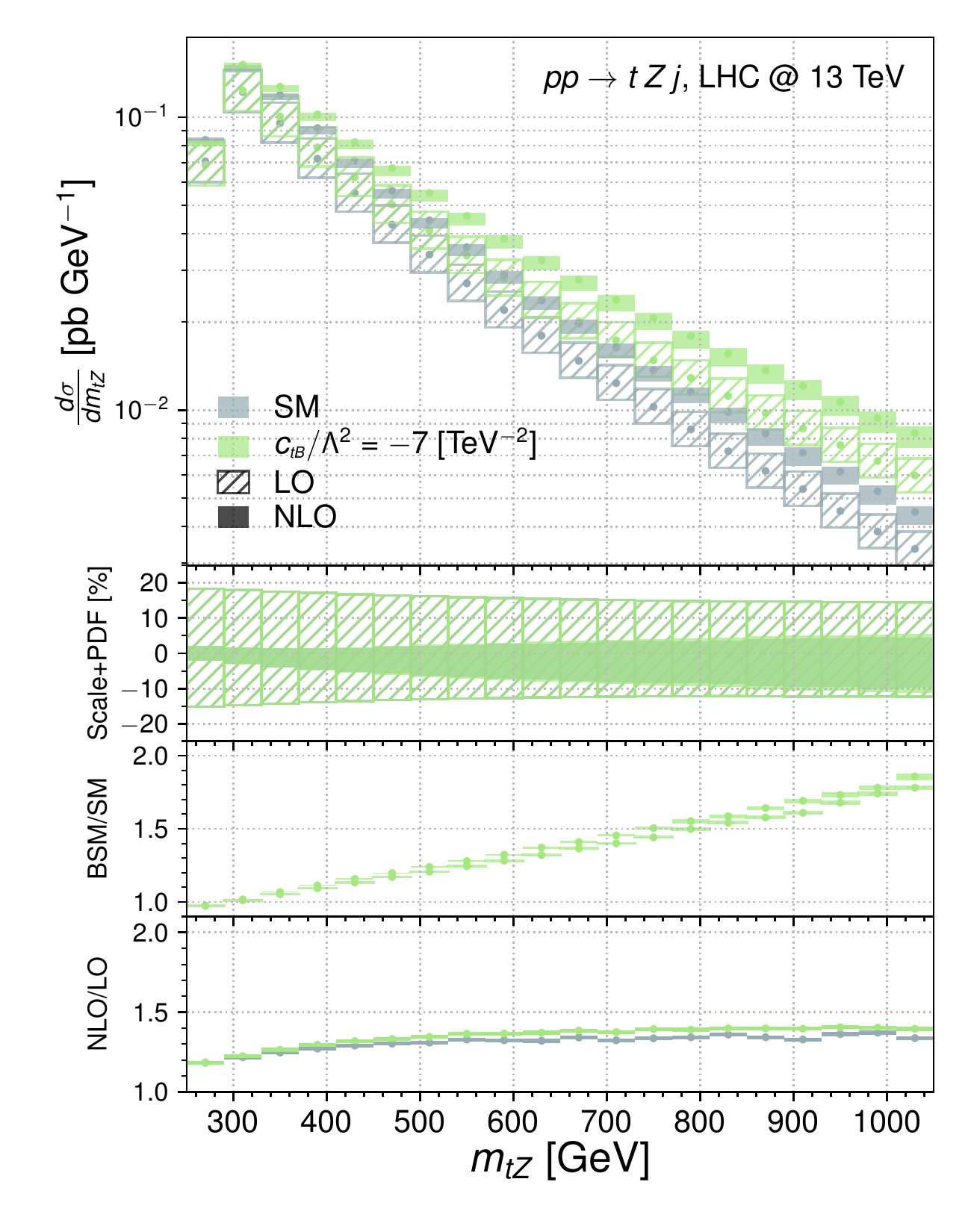}\\
    \includegraphics[width=0.45\linewidth]{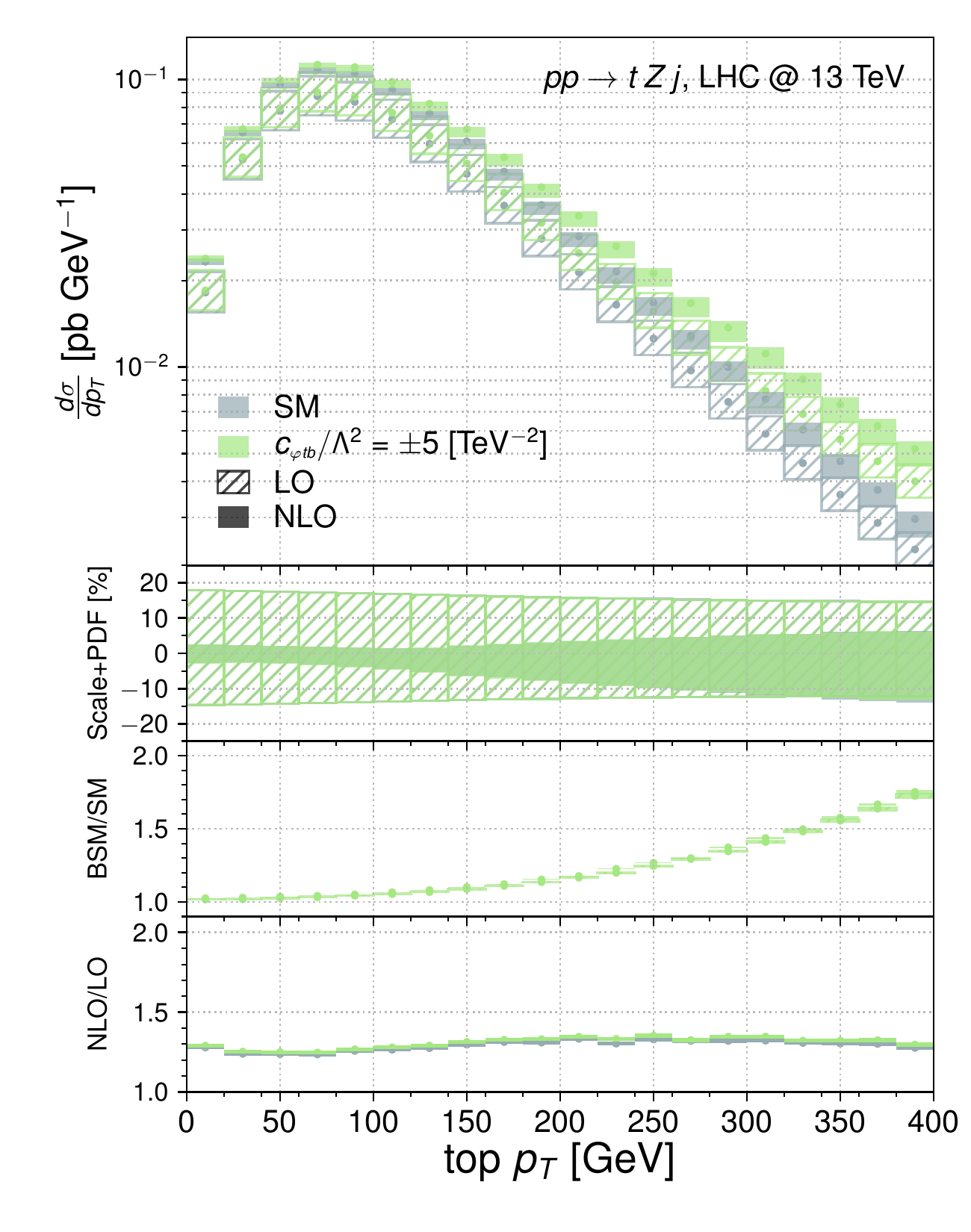}
    \includegraphics[width=0.45\linewidth]{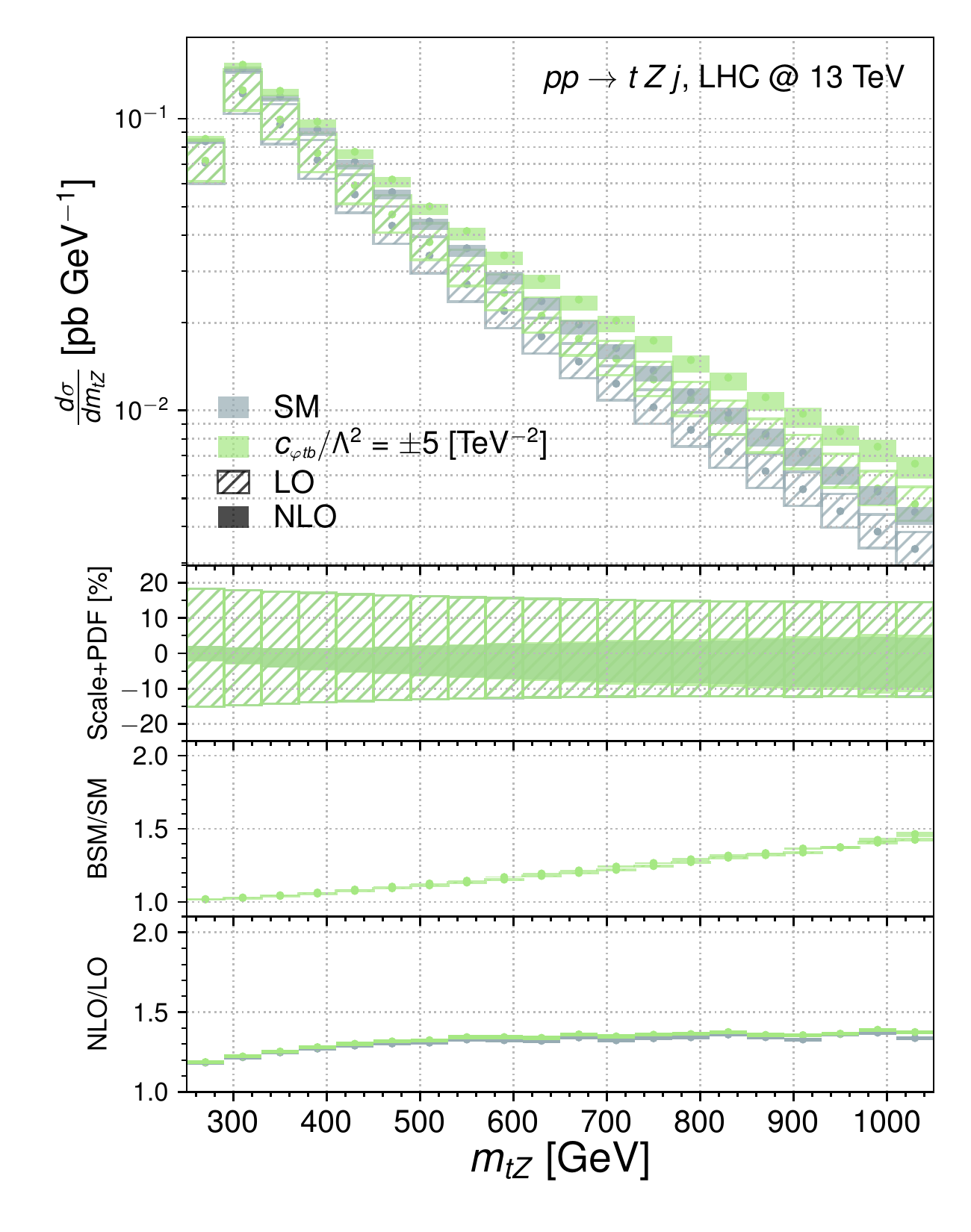}
    \caption{\label{fig:tzj_distributions}
    Differential distributions of the top $p_T$ and top-$Z$ system invariant mass 
    for the $tZj$ process for given values of the $\Op{tB}$  and $\Op{\phi tb}$ operator coefficients roughly saturating current individual, direct limits. The lower insets 
    show the scale and PDF uncertainty bands, the ratio over the SM prediction and finally 
    the corresponding $K$-factor.  }

\end{figure}

A selection of distributions are shown in Figures~\ref{fig:thj_distributions} and \ref{fig:tzj_distributions}. The already large effects at inclusive level are amplified in the tails of the $p_T$ distributions, with significant energy growth present in all distributions shown. The $tHj$ deviations reach factors of many in the tails, while for $tZj$, the 20\% inclusive effects become a factor of a few in the high energy bins. There is therefore a complementarity between the two processes since, although the largest effects are present in $tHj$, the process is comparatively rare and may not be probed differentially at the LHC, at least until the late high-luminosity phase. $tZj$, however has  a ten times larger cross section and could therefore gather enough statistics for differential measurements and an enhanced sensitivity to the operators in question.

\subsection{Current and future sensitivity}
\label{sec:sensitivity}
The two most recent measurements of the $tZj$ process~\cite{Aaboud:2017ylb,Sirunyan:2017nbr} allow for a first sensitivity assessment of this process to the EFT coefficients of interest at the inclusive level. The experiments perform fits to the signal strength, $\mu$, with respect to the SM expectation in this channel to extract the measured cross section. In order to eliminate some dependence on the overall normalisation and reduce scale uncertainties, we construct confidence intervals on the Wilson coefficients by performing a $\Delta\chi^2$ fit to the signal strength directly rather than the measured cross section. The ratio of the $tZj$ cross section over the SM one as a function of the Wilson coefficients is taken from the results of Table~\ref{tZj:sigma} and compared to the observed values of $\mu=0.75\pm0.27$ and $1.31\pm0.47$ reported by CMS and ATLAS respectively, where the uncertainty is taken to be the sum in quadrature of the statistical and systematic components. Both measurements are made searching for the electron and muon decay modes of the $Z$-boson on-shell, {\it i.e.}, including a cut on the dilepton invariant mass. We therefore take into account the modification of these branching fractions in the presence of the $\Op{\phi Q}^{\sss (1)}$ and $\Op{\phi Q}^{\sss (3)}$ operators.

\begin{figure}
    \centering
    \subfloat[Current $tZj$]{
    \includegraphics[width=0.35\linewidth]{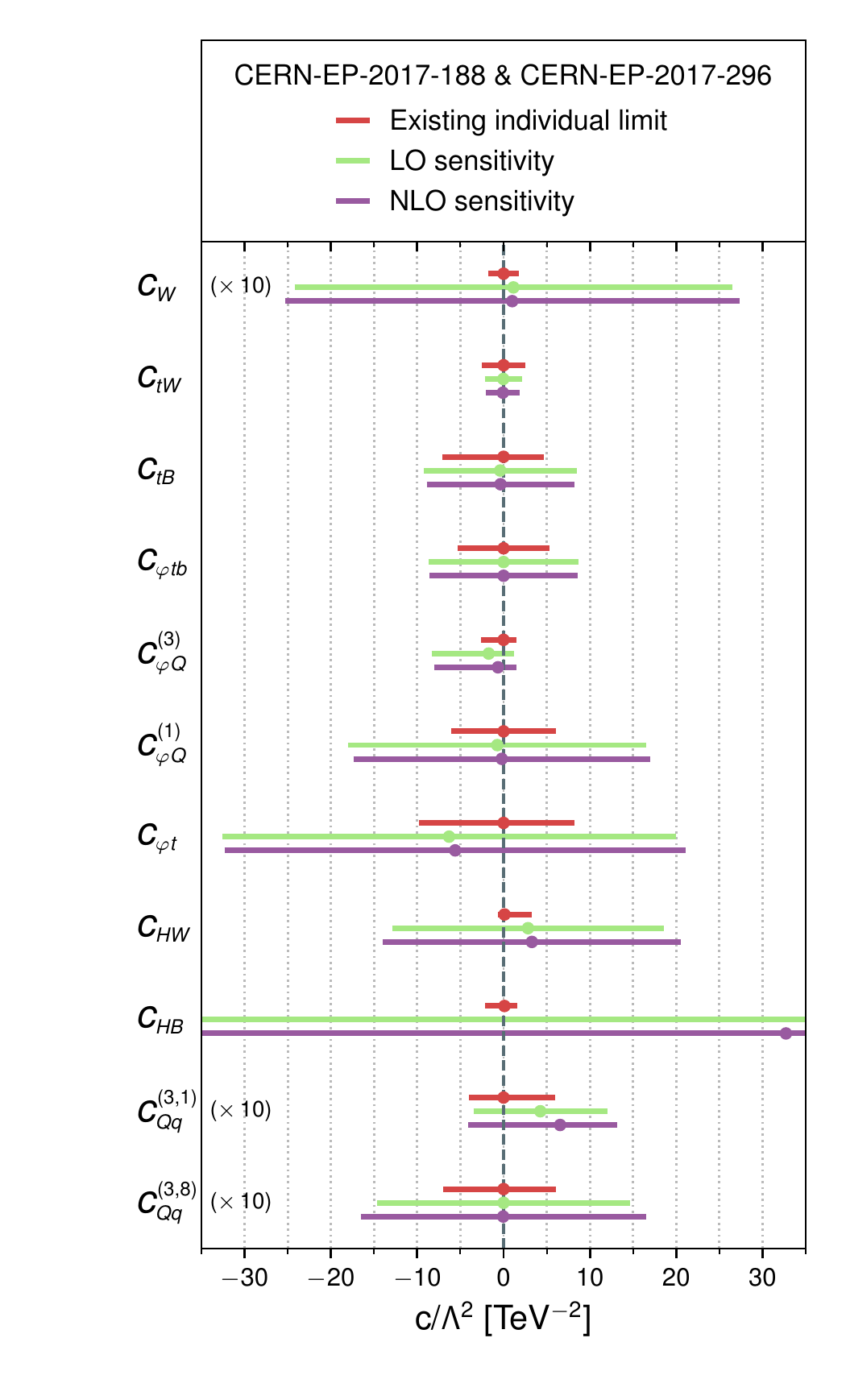}
    }
    \subfloat[Current $tHj$]{
    \includegraphics[width=0.35\linewidth]{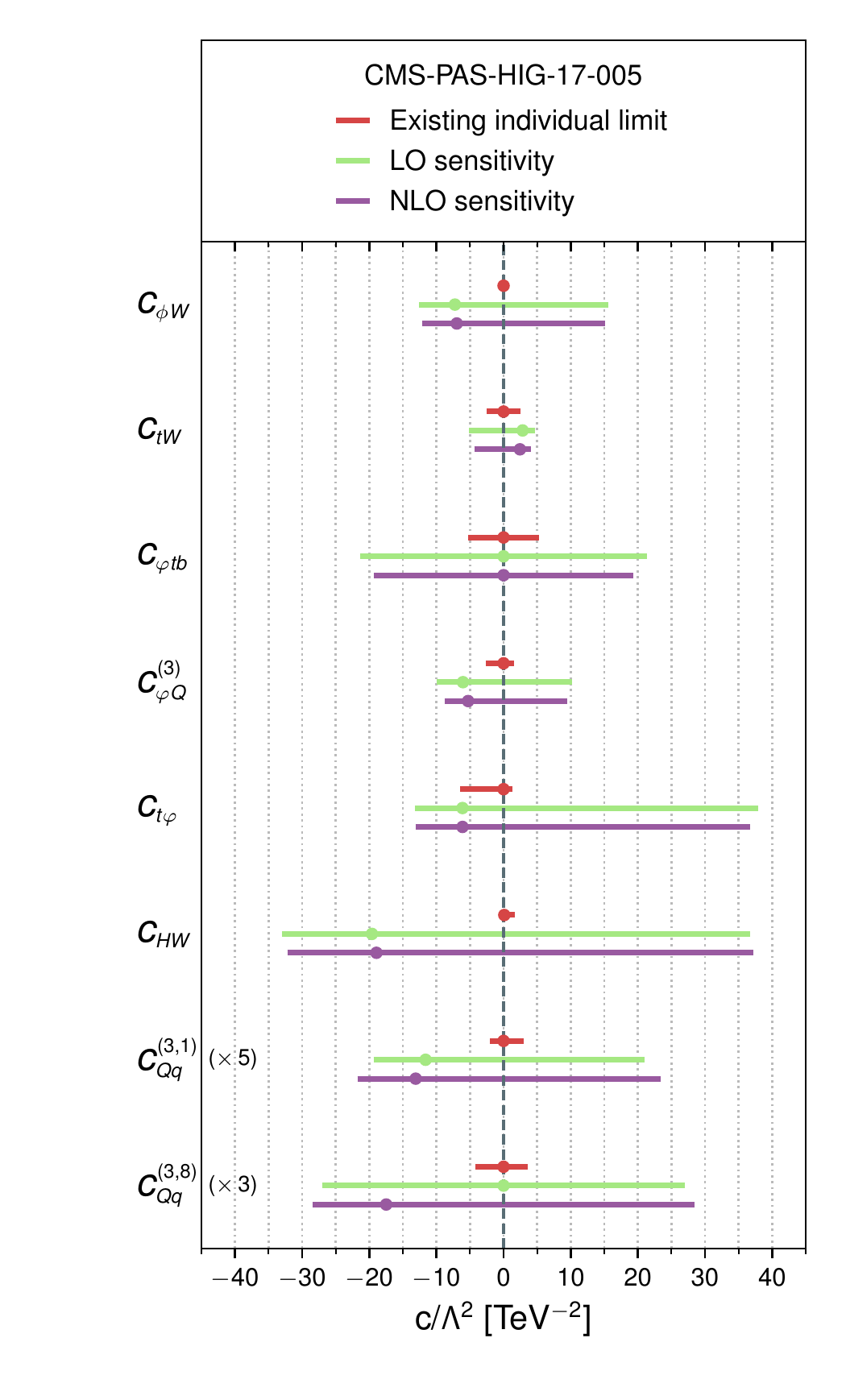}
    }\\
    \subfloat[Future, high $p_T$ $tZj$]{
    \includegraphics[width=0.35\linewidth]{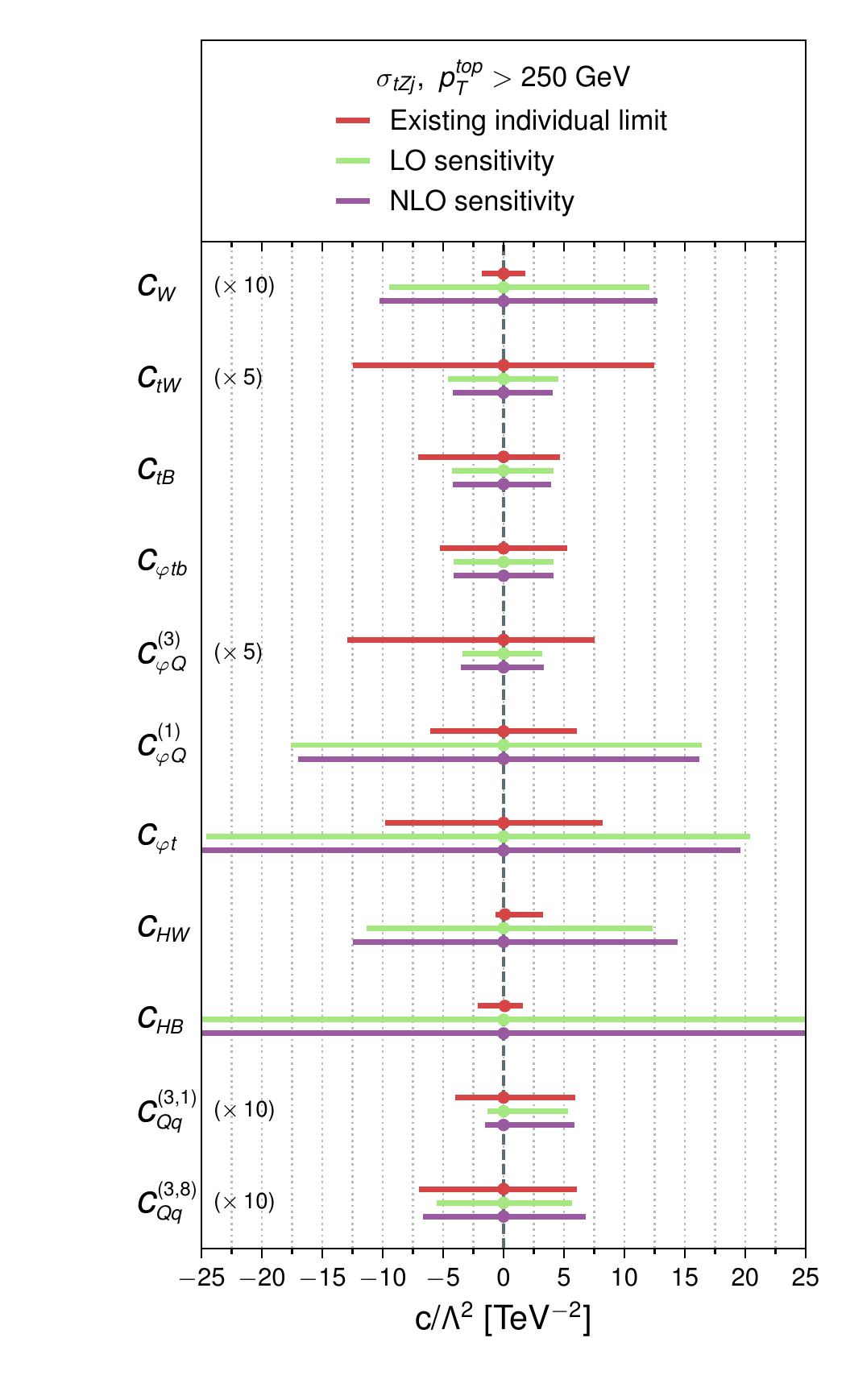}
    }
    \subfloat[Future $tHj$]{
    \includegraphics[width=0.35\linewidth]{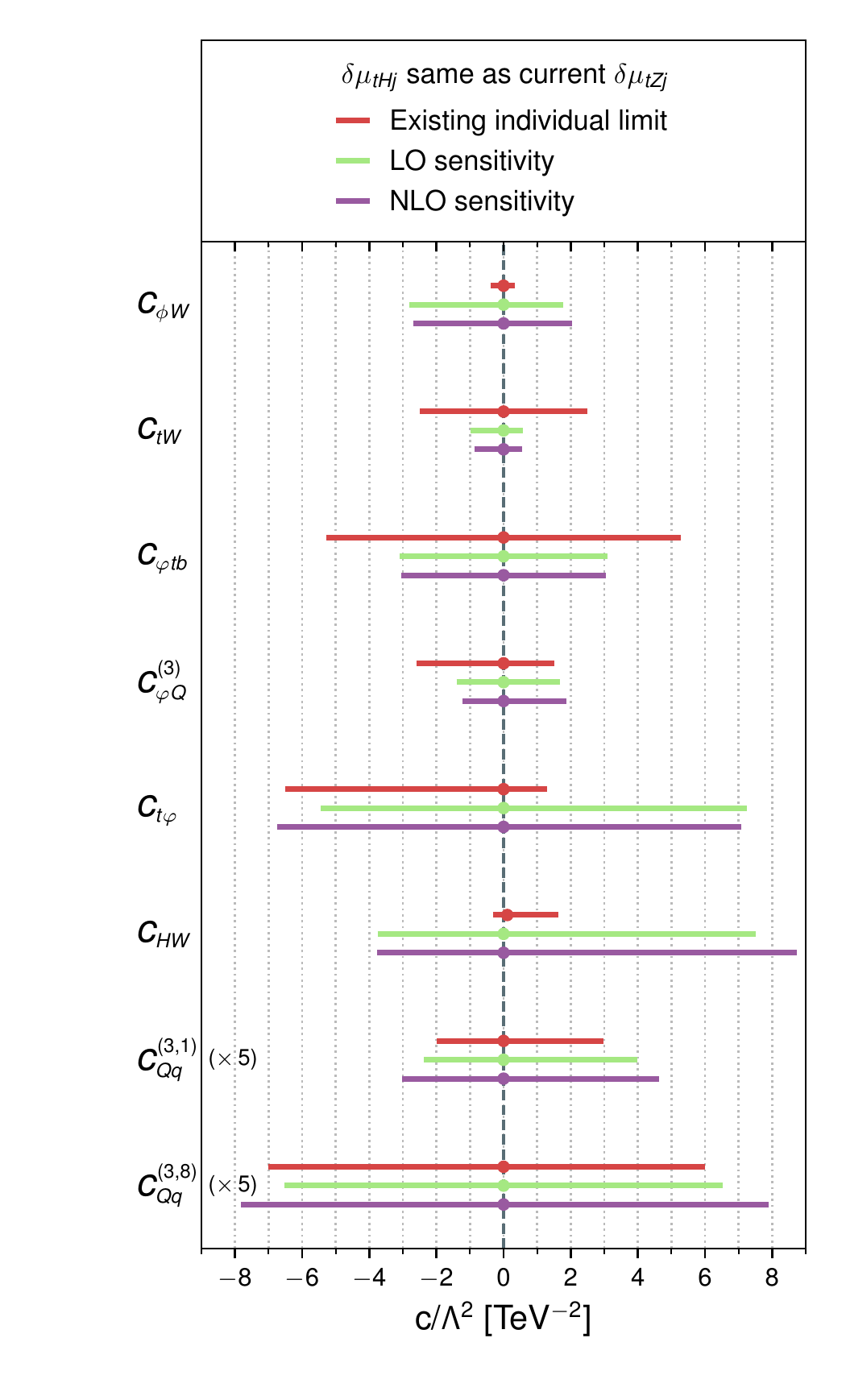}
    }
    \caption{\label{fig:tzj_sensitivity}
    Confidence intervals on the Wilson coefficients of interest derived from  a) the signal strength measurements of the $tZj$ process by ATLAS~\cite{Aaboud:2017ylb} and CMS~\cite{Sirunyan:2017nbr} and b) from the combined signal strength measurement of $t\bar{t}H$ $tHj$ and $tHW$ by CMS~\cite{CMS:2017uzk}  assuming only modifications to $tHj$ apart from the modifications to $t\bar{t}H$ induced by $\Op{t\phi}$. Future sensitivity is shown in c) for $tZj$, assuming the same accuracy as for the current inclusive measurement is achieved in the $p^{top}_T > 250$ GeV region and in d) for $tHj$, assuming the accuracy of the current $tZj$ inclusive measurement is achieved. See text for further details. In all cases the existing limits quoted in Table~\ref{tab:constraints} as also included for reference.}
\end{figure}

Note that this procedure is rather simplistic and uncertain given the complexity of the $tZj$ measurement the LHC. Firstly, due to the relatively small rates and large potential background contributions, multivariate analysis methods are employed to improve the signal to background ratio. The efficiency and acceptance factors that are used in the extrapolation to the full phase space apply strictly to the SM kinematics and may be different in general for the EFT. One is only truly sensitive to enhancements of the cross section in the observed fiducial region after selection requirements. Furthermore, the signal yields are fitted using templates for the multivariate classifier output, which may also differ between the SM and EFT. Finally, many of the backgrounds considered in this analysis would also be affected non-negligibly by the presence of the same operators. The dominant di-boson background, for example would be modified by $\Op{W}$ while several others, such as $t\bar{t}V$, $t\bar{t}H$ and $tWZ$ would get affected by a combination of top and EW operators. Our confidence intervals are obtained neglecting all of these effects and should therefore be viewed as approximate sensitivity estimates. Figure~\ref{fig:tzj_sensitivity}~(a) reports the obtained confidence intervals compared to the existing individual limits from Table~\ref{tab:constraints}. In most cases, the current inclusive measurement does not probe the operators beyond existing limits. The single exception is in the case of the weak dipole operator, $\Op{tW}$. The enhanced relative squared dependence on this operator  leads to a slightly improved sensitivity over the individual limit obtained from a combination of LHC Run 1 single-top and $W$ helicity fraction measurements. 
%

The differential results of Section~\ref{sec:differential} indicate that more information may be provided by a future measurement of this process, particularly at high $p_T$. In order to test this, we consider a hypothetical future measurement of $tZj$ in the high energy region, in which the top transverse momentum is required to be above 250 GeV. In the SM, the predicted cross section at NLO in this phase space region is 69 fb, roughly a factor 10 smaller compared to the inclusive prediction. Remaining agnostic about the nature of a future analysis, we assume that such a cross section should be attainable with the same precision as the current measurement with about 10 times more data. This suggests that one could expect this level of sensitivity in the early stages of the high-luminosity LHC run. Our projected sensitivities, shown in Figure~\ref{fig:tzj_sensitivity}~(c), are obtained assuming the SM prediction, $\mu=1$,  observed by both experiments and taking the same uncertainties as for the inclusive measurement. As expected, we see significant improvements, particularly for $\Op{tW}$, $\Op{tB}$, $\Op{\phi Q}^{\sss (3)}$, $\Op{\phi tb}$, $\Op{Qq}^{\sss   (3,1)}$ and $\Op{Qq}^{\sss   (3,8)}$, that may reach beyond the current limits summarised in Table~\ref{tab:constraints}. Considering the high energy growth of the sub amplitudes of Table~\ref{tab:bwtz}, one can see that the large relative gains in sensitivity all occur for operators with the strongest energy growths while for operators without many enhanced helicity configurations such as $\Op{\phi t}$ or $\Op{\phi Q}^{\sss (1)}$ do not benefit at all. We note 
in particular the improvement on the limit on $\Op{\phi Q}^{\sss (3)}$ due to the unitarity violating behaviour of the
amplitude at high-energy, a feature not present in single top production where $\Op{\phi Q}^{\sss (3)}$ uniformly 
rescales the cross section, as discussed in Section~\ref{sec:energybehaviour}. Although the four-fermion operators can be constrained significantly better than from Run 1 single top, we expect that forthcoming Run 2 single-top measurements will constrain such operators better.


As of today, the $tHj$ process has yet to be measured in isolation at the LHC. However, several searches have been performed in which this process is a part of the signal selection~\cite{CMS-PAS-HIG-16-019,CMS:2017uzk}. The former sets an upper limit of 113 times the SM prediction on the combination of $tHj$ and $tHW$ processes with 2.3 fb$^{-1}$ of integrated luminosity while the latter additionally includes the $t\bar{t}H$ process and obtains a combined signal strength for the SM hypothesis of $\mu=1.8\pm0.67$ with 35.9 fb$^{-1}$. Since the former analysis lacks sensitivity due to the small dataset used, we use the second measurement to estimate current sensitivity to the $tHj$ process, accepting a large amount of pollution from $t\bar{t}H$. In this case we assume that only the $tHj$ process is modified apart from the contribution to $t\bar{t}H$ from the top Yukawa operator, $\Op{t\phi}$, obtained from~\cite{Maltoni:2016yxb}. This operator affects the dominant, QCD-induced component of $t\bar{t}H$, while the other operators that we consider would only contribute to the EW component, which in the SM is more than two orders of magnitude below the QCD one. Similarly, the $tHW$ process is about five times smaller than $tHj$ in the SM. Furthermore, since the measurement targets the $\tau\tau$, $WW$ and $ZZ$ decay modes of the Higgs, we also take into account the effect of the modified branching ratios due to $\Op{\phi W}$ at LO. The sensitivity estimates from this measurement are shown in Figure~\ref{fig:tzj_sensitivity}~(b), and suggest that a significant improvement is needed to obtain relevant constraints on the operators of interest. 

Phenomenological studies on future $tHj$ prospects in the SM have been performed for the high-luminosity LHC run~\cite{Biswas:2012bd,Chang:2014rfa}, concluding that it may be possible to access this mode with the full design integrated luminosity of 3 ab$^{-1}$. For our purposes, we consider the possibly optimistic scenario in which the process is measured with the same sensitivity as the current $tZj$ measurement, just to highlight the gain that would occur in this hypothetical case. Figure~\ref{fig:tzj_sensitivity}~(d) clearly shows a marked improvement. In the case of the dipole and RHCC, the potential sensitivity goes beyond that of the high-$p_T$ $tZj$, while for the four-fermion operators, the benefit of looking at the kinematic tails of $tZj$ outweighs the strong dependence of the inclusive $tHj$ cross section.
%

Overall, the interesting individual sensitivity prospects concerning the operators included in our study mainly cover the weak dipole, RHCC and single-top four-fermion operators, with the sensitivity to most of the current-current, triple gauge and gauge-Higgs operators remaining below the existing limits from other measurements of less rare and already established processes such as single-top, diboson and Higgs production/decay. The main exception to this is with $\Op{\phi Q}^{\sss(3)}$, for which a new, interfering, energy growth arises and will lead to significant improvement on current sensitivities through high energy $tZj$ measurements. Nevertheless, when performing a global analysis and marginalising over the various operators, these processes may well provide some additional constraining power also in these directions towards the latter stages of the LHC lifetime. 

\section{Conclusions}

Electroweak production of a single top quark in association with a $Z$ or Higgs boson provides a natural opportunity to constrain possible deviations of the neutral couplings of the top quark with respect to the SM predictions. The motivations and interest for this class of processes are multifold. First, being mediated only by electroweak interactions at LO, they can be predicted accurately in perturbative QCD, already at NLO accuracy, and they are not affected by possible deviations in the QCD interactions (at LO).  Second, these processes feature an enhanced sensitivity, appearing as a non-trivial energy dependence, also for operators that, per se, do not necessarily lead to interactions that grow with energy. This is due to the spoiling of delicate gauge cancellations that take place in the SM, when anomalous interactions are present. Last, but not least, these processes are of phenomenological interest, as they are already being studied at the LHC. 

In this work we have considered for the first time $tHj$ and $tZj$ in the context of the standard model effective field theory, in the presence of all the relevant dim-6 operators. We have included NLO QCD corrections and studied the relevant theoretical uncertainties on our predictions. As expected, while not very large in general, QCD corrections typically reduce the theoretical uncertainties and can lead to non-flat $K$-factors for differential observables. Using the measurements of the signal strengths of these processes at the LHC we have performed a  first sensitivity study allowing one non-zero operator coefficient at a time. This study can be therefore considered the first necessary step before performing a global fit. Whilst at the moment the constraints from $tZj$ measurements cannot compete with the already existing limits on the operators of interest, there is enough evidence that complementary constraints could be obtained within the projected experimental accuracies. 

Given the promising signs found already at the inclusive level, we have examined the impact of the dim-6 operators  on differential observables such as the top-quark transverse momentum and the invariant mass of the top-quark-$H/Z$  system. We have found that the effects on the total cross section are typically amplified at the tails of 
distributions leading to allowed deviations from the SM predictions of a factor of a few. We have argued that this behaviour is directly related to the energy behaviour of the relevant sub-amplitudes $b\,W\to t\,h$ and $b\,W\to t\,Z$ involved in $tHj$  and $tZj$, respectively, which we have also reported in detail. New sources of energy growth not present in, {\it e.g.}, single top production are identified and exploited in our sensitivity studies.

Our findings support extracting useful constraints from inclusive and/or differential measurements of the $tHj$ and $tZj$ processes, which are expected at the high-luminosity LHC.  For example, given the current constraints on the weak dipole and right handed charged current operators, very large deviations can be still expected in both $tZj$ and $tHj$. In addition, the information that could be extracted on the Yukawa operator could also become competitive with enough integrated luminosity. Whilst not discussed in this work, we have also verified that $t \gamma j$  displays similar sensitivities as $tZj$ to the same class of dim-6 operators. A dedicated study of this process with the goal of motivating a measurement at the LHC, which to our knowledge is not being pursued yet, is ongoing.

In summary, we have proposed to use measurements of  $tZj$ and $tHj$ at the LHC to constrain the least known operators in the SMEFT, {\it i.e.}, those involving top-quark, gauge and Higgs interactions. We have computed  $tHj$ and $tZj$ cross sections in the SMEFT at NLO in QCD, achieving for the first time such an accuracy  for processes where the three types of operators, namely, purely gauge, two-fermion and four-fermion operators, can contribute.  This work proves that it is now possible to obtain NLO accurate predictions automatically for any dim-6 operator  and process involving top-quarks, weak bosons and Higgs final states and therefore paves the way to performing global SMEFT fits at the LHC. 

\acknowledgments
We would like to thank Liam Moore and Ambresh Shivaji for discussions. FM has received fundings from the European Union's Horizon 2020 research and innovation programme as part of the Marie Sk\l{}odowska-Curie Innovative Training Network MCnetITN3 (grant agreement no. 722104) and by the  F.R.S.-FNRS under the `Excellence of Science` EOS be.h project n. 30820817. EV and KM are supported by a Marie Sk\l{}odowska-Curie Individual Fellowship of the European Commission's Horizon 2020 Programme under contract numbers 704187 and 707983, respectively.  CZ is supported by IHEP under Contract No. Y7515540U1. Computational resources have been provided by the supercomputing facilities of the Universit\'e catholique de Louvain (CISM/UCL) and the Consortium des \'Equipements de Calcul Intensif en F\'ed\'eration Wallonie Bruxelles (C\'ECI).

\begin{landscape}
  \begin{table}[h!]
      \centering
     \renewcommand{\arraystretch}{1.}
\begin{center}
\begin{tabular}{|c|c|cccccc|cc|}
    \hline
 ${\sst \lambda_b}$, ${\sst \lambda_W}$, ${\sst \lambda_t}$, ${\sst \lambda_Z}$ & SM  & $\Op{\phi Q}^{\sss (3)}$ & $\Op{\phi Q}^{\sss (1)}$ & $\Op{\phi t}$& $\Op{tB}$ & $\Op{tW}$ & $\Op{W}$&$\Op{HW}$&$\Op{HB}$\tabularnewline
\hline
$-, 0, -, 0$& 
$s^0$ & ${\sst \sqrt{s (s+t)}}$ & $-$ & $-$ & $-$ & $s^0$ & $s^0$ &${\sst \sqrt{s (s+t)}}$ & $s^0$
\tabularnewline
$-, 0, +, 0$& 
$\frac{1}{\sqrt{s}}$ & ${\sst m_t\sqrt{-t}}$ & ${\sst m_t\sqrt{-t}}$ & $ {\sst m_t\sqrt{-t}}$  & $ {\sst m_Z\sqrt{-t}}$ & $\frac{ m_W (2s+3t)}{ \sqrt{-t} }$ & $-$ &$ {\sst m_t\sqrt{-t}}$&$ {\sst m_t\sqrt{-t}}$
\tabularnewline
$-, -, -, 0$& 
$\frac{1}{\sqrt{s}}$ & $ {\sst m_W\sqrt{-t}}$ & $-$ & $-$ &  $-$ & $-$ & $\frac{m_W (s+2 t)}{\sqrt{-t}}$ & $ {\sst m_W\sqrt{-t}}$ &$\frac{1}{\sqrt{s}}$
\tabularnewline
$-, -, +, 0$& 
$\frac{1}{s}$  & $s^0$ & $s^0$ & $s^0$  & $s^0$ & ${\sst \sqrt{s (s+t)}}$ & $s^0$ &$s^0$ &$\frac{1}{\sqrt{s}}$
\tabularnewline
$-, 0, -, -$& 
$\frac{1}{\sqrt{s}}$ & ${\sst m_W\sqrt{-t}} $ & $-$ & $-$ & $ {\sst m_t\sqrt{-t}}$ & ${\sst m_t\sqrt{-t}}$ & $\frac{m_W (s+2 t)}{\sqrt{-t}}$ &$\frac{m_W (s s_{W}^2 +2 t)}{\sqrt{-t}}$ & $\frac{m_W\,s}{\sqrt{-t}}$
\tabularnewline
$-, 0, -, +$& 
$\frac{1}{\sqrt{s}}$  & $-$ & $-$ & $-$  & $-$ & $-$ & $\frac{m_W (s+t)}{\sqrt{-t}}$ &$\frac{m_W (s+t)}{\sqrt{-t}}$
&$\frac{m_W (s+t)}{\sqrt{-t}}$
\tabularnewline
$-, 0, +, -$& 
$s^0$ & $s^0$ & $s^0$ & $-$  & $-$ & $s^0$ & $s^0$ &$s^0$&$s^0$
\tabularnewline
$-, 0, +, +$& 
$\frac{1}{s}$ & $s^0$ & $s^0$ & $s^0$ &  ${\sst \sqrt{s (s+t)}}$ & ${\sst \sqrt{s (s+t)}}$ & $-$ &$s^0$&$s^0$
\tabularnewline
$-, +, -, 0$& 
$\frac{1}{\sqrt{s}}$ & $-$ & $-$ & $-$  & $-$ & $-$ & $\frac{ m_W (s+t)}{\sqrt{-t}}$ &$\frac{1}{\sqrt{s}}$&$\frac{1}{\sqrt{s}}$
\tabularnewline
$-, +, +, 0$& 
$s^0$  & $s^0$ & $-$ & $-$ & $-$ & $s^0$ & $-$ &$s^0$&$\frac{1}{s}$
\tabularnewline
$-, -, -, -$& 
$s^0$  & $s^0$ & $s^0$ & $-$ & $s^0$ & $s^0$ & $s^0$&$s^0$&$s^0$
\tabularnewline
$-, -, -, +$& 
$\frac{1}{s}$  & $-$ & $-$ & $-$ & $-$ & $-$ & ${\sst \sqrt{s (s+t)}}$ &$s^0$&$s^0$
\tabularnewline
$-, -, +, -$& 
$\frac{1}{\sqrt{s}}$ & $-$ & $-$ & $-$  & $-$ & $\frac{m_Z \left( s_{W}^2 t-3 c_{W}^2 (2 s+t)\right)}{ \sqrt{-t}}$ & $-$&$\frac{1}{\sqrt{s}}$&$\frac{1}{\sqrt{s}}$
\tabularnewline
$-, -, +, +$& 
$-$ & $-$ & $-$ & $-$ & $ {\sst m_W\sqrt{-t}}$ & $ {\sst m_Z \sqrt{-t} }$ & ${\sst m_t\sqrt{-t}} $ & ${\sst m_t\sqrt{-t}} $&${\sst m_t\sqrt{-t}} $
\tabularnewline
$-, +, -, -$& 
$\frac{1}{s}$  & $-$ & $-$ & $-$  & $-$ & $-$ & ${\sst \sqrt{s (s+t)}}$ & $s^0$ & $s^0$ 
\tabularnewline
$-, +, -, +$& 
$s^0$ & $s^0$ & $s^0$ & $-$  & $-$ & $-$ & $-$ &  $s^0$ & $s^0$ 
\tabularnewline
$-, +, +, -$& 
$\frac{1}{\sqrt{s}}$  & $-$ & $-$ & $-$ & $-$ & $-$ & ${\sst m_t\sqrt{-t}} $ & ${\sst m_t\sqrt{-t}} $&${\sst m_t\sqrt{-t}} $
\tabularnewline
$-, +, +, +$& 
$\frac{1}{\sqrt{s}}$  & $-$ & $-$ & $-$ & $-$ & $\frac{m_W (s+t)}{\sqrt{-t}}$ & $-$ & $\frac{1}{\sqrt{s}}$&$\frac{1}{\sqrt{s}}$
\tabularnewline
\hline
\end{tabular}
\vspace{1cm}
\renewcommand{\arraystretch}{1.}
\begin{minipage}{0.45\linewidth}
\vspace{0.5cm}
\centering
$\Op{\phi tb},\,\lambda_b,\lambda_t=+,+$\\
\begin{tabular}{|c|ccc|}
\hline
\diagbox[height=0.8cm,innerwidth=0.8cm,font=\footnotesize]{${\sst\lambda_Z}$}{${\sst\lambda_{\sss W}}$}&0&$+$&$-$\tabularnewline
\hline
\renewcommand{\arraystretch}{1.}
$0$& ${\sst \sqrt{s (s+t)}}$ & ${\sst m_W\sqrt{-t}}$ &  $-$  \tabularnewline
$+$& ${\sst m_Z\sqrt{-t}}$ & $s^0$   & $-$ \tabularnewline
$-$& $-$ & $-$   & $s^0$ \tabularnewline
\hline
\end{tabular}
\end{minipage}
\renewcommand{\arraystretch}{1.}
\begin{minipage}{0.45\linewidth}
\vspace{0.5cm}
\centering
$\Op{\phi tb},\,\lambda_b,\lambda_t=+,-$\\
\begin{tabular}{|c|ccc|}
\hline
\diagbox[height=0.8cm,innerwidth=0.8cm,font=\footnotesize]{${\sst\lambda_Z}$}{${\sst\lambda_{\sss W}}$}&0&$+$&$-$\tabularnewline
\hline
\renewcommand{\arraystretch}{1.4}
$0$& $-$ & $-$ & $s^0$  \tabularnewline
$+$& $s^0$ & $-$   & $-$ \tabularnewline
$-$& $s^0$ & $-$   & $-$ \tabularnewline
\hline
\end{tabular}
\end{minipage}
%
\vspace{-1cm}
\end{center}

      \caption{\label{tab:bwtz}
      Energy growth for helicity amplitudes in the $b\,W\to t\,Z$ subamplitude 
      in the high energy limit, $s , -t \gg v$ with $s/-t$ constant. The RHCC operator contributions are collected 
    separately due to the fact that it is the only one that can yield right 
    handed $b$-quark configurations in the 5-flavour scheme.
      }
  \end{table} 
\end{landscape}
\clearpage
\begin{landscape}
\begin{table}[ht!]
\centering
\renewcommand{\arraystretch}{1.}
\centering
\small
\begin{tabular}{l|lllllll}
$\sigma_{ij}$&$c_{{\sss \phi W}}$&$c_{{\sss t\varphi}}$&$c_{{\sss tW}}$&$c^{\sss (3)}_{{\sss \varphi Q}}$&$c_{{\sss HW}}$&$c_{{\sss u31}}$&$c_{{\sss u38}}$\tabularnewline
\hline
$c_{{\sss \phi W}}$&$-$&2.752 (1.29)&12.88 (0.61)&6.384 (0.65)&-0.43 (-0.17)&$-$&$-$\tabularnewline
$c_{{\sss t\varphi}}$&2.514 (1.35)&$-$&-1.912 (-0.27)&-4.168 (-1.25)&-0.699 (-0.80)&$-$&$-$\tabularnewline
$c_{{\sss tW}}$&10.54 (0.68)&-1.772 (-0.32)&$-$&-26.24 (-0.79)&3.988 (0.46)&$-$&$-$\tabularnewline
$c^{\sss (3)}_{{\sss \varphi Q}}$&5.12 (0.67)&-3.584 (-1.31)&-11.2 (-0.49)&$-$&4.864 (1.21)&$-$&$-$\tabularnewline
$c_{{\sss HW}}$&-0.402 (-0.18)&-0.6138 (-0.78)&3.124 (0.47)&3.5784 (1.10)&$-$&$-$&$-$\tabularnewline
$c_{{\sss u31}}$&-13.475 (-0.71)&5.16 (0.76)&-19.1 (-0.34)&-15.44 (-0.55)&-6.96 (-0.86)&$-$&4.525 (0.15)\tabularnewline
$c_{{\sss u38}}$&$-$&$-$&$-$&$-$&$-$&$-$&$-$\tabularnewline
\end{tabular}
  \caption{Cross-section results for $tHj$ at 13 TeV, following the parametrisation of Eq.~ (\ref{eq:xsecpara}) at LO (lower left half) and NLO (upper left half). Central values for the EFT cross-terms are quoted in fb followed by their ratio to the geometric mean of the two squared terms of the respective operators.}
        \label{tHj:sigma_crs}
\end{table}
\begin{table}[ht!]
\centering
\renewcommand{\arraystretch}{1.}
\setlength\tabcolsep{0.1cm}
\centering
\small
\begin{tabular}{l|llllllllll}
$\sigma_{ij}$&$c_{{\sss W}}$&$c_{{\sss tW}}$&$c_{{\sss tB}}$&$c^{\sss (3)}_{{\sss \varphi Q}}$&$c_{{\sss \varphi Q}}$&$c_{{\sss tR}}$&$c_{{\sss HW}}$&$c_{{\sss HB}}$&$c_{{\sss u31}}$&$c_{{\sss u38}}$\tabularnewline
\hline
$c_{{\sss W}}$&$-$&-3.28 (-0.11)&-0.62 (-0.05)&-2.92 (-0.36)&0.5 (0.07)&-0.2 (-0.05)&-1.16 (-0.65)&$-$&$-$&$-$\tabularnewline
$c_{{\sss tW}}$&-2.58 (-0.13)&$-$&5.46 (0.60)&3.12 (0.57)&-0.368 (-0.08)&0.35 (0.12)&0.836 (0.69)&$-$&$-$&$-$\tabularnewline
$c_{{\sss tB}}$&-0.36 (-0.04)&3.6 (0.62)&$-$&-0.056 (-0.02)&-0.549 (-0.27)&0.712 (0.53)&0.238 (0.43)&-0.025 (-0.14)&$-$&$-$\tabularnewline
$c^{\sss (3)}_{{\sss \varphi Q}}$&2.2 (0.35)&1.984 (0.54)&-0.108 (-0.06)&$-$&2.068 (1.67)&0.112 (0.14)&3.232 (9.80)&-0.536 (-4.83)&$-$&$-$\tabularnewline
$c_{{\sss \varphi Q}}$&0.524 (0.10)&-0.438 (-0.15)&-0.51 (-0.35)&1.456 (1.60)&$-$&-0.831 (-1.24)&-0.456 (-1.67)&0.0517 (0.56)&$-$&$-$\tabularnewline
$c_{{\sss tR}}$&-0.207 (-0.06)&0.246 (0.12)&0.572 (0.59)&0.296 (0.49)&-0.6561 (-1.32)&$-$&0.286 (1.60)&-0.0287 (-0.48)&$-$&$-$\tabularnewline
$c_{{\sss HW}}$&-0.9 (-0.57)&0.568 (0.61)&0.2 (0.44)&2.2 (7.73)&-0.44 (-1.89)&0.284 (1.83)&$-$&-0.0762 (-3.12)&$-$&$-$\tabularnewline
$c_{{\sss HB}}$&$-$&$-$&-0.023 (-0.17)&-0.4364 (-5.00)&0.0413 (0.58)&-0.0218 (-0.46)&-0.062 (-2.78)&$-$&$-$&$-$\tabularnewline
$c_{{\sss u31}}$&81.5 (0.32)&-16.1 (-0.11)&-1.6 (-0.023)&-25.3 (-0.57)&3.8 (0.11)&-8.3 (-0.34)&-9.5 (-0.46)&3.7 (0.58)&$-$&$-$\tabularnewline
$c_{{\sss u38}}$&$-$&$-$&$-$&$-$&$-$&$-$&$-$&$-$&$-$&$-$\tabularnewline
\end{tabular}
  \caption{Cross-section results for $tZj$ at 13 TeV, following the parametrisation of Eq.~(\ref{eq:xsecpara}) at LO (lower left half) and NLO (upper left half). Central values are quoted in fb followed by the ratio of the interference terms and the geometric mean of the two squared terms of the respective operators.}
        \label{tZj:sigma_crs}
\end{table}
\end{landscape}

\bibliography{refs.bib}

\end{document}